\documentclass[nofootinbib,twocolumn,showpacs,pre,aps,superscriptaddress]{revtex4-2}

\usepackage[dvipdfmx]{graphicx}
\usepackage{bm}
\usepackage{amsmath}
\usepackage{amssymb}
\usepackage{newtxtext}
\usepackage{newtxmath}
\usepackage{color}
\usepackage{tikz}
\usepackage{esint}
\usepackage{float} 
\allowdisplaybreaks[1]

\usepackage{color}

\begin{document}


\title{Lateral hydrodynamics in supported membranes: \\The Evans-Sackmann model and its extensions}

\author{Yuto Hosaka}\email{hosaka.yuto.7r@kyoto-u.ac.jp\\
ORCID: 0000-0002-6202-4206}
\affiliation{Department of Mathematics, Kyoto University, Kyoto 606-8502, Japan}

\author{David Andelman}\email{andelman@tauex.tau.ac.il\\
ORCID: 0000-0003-3185-8475}
\affiliation{School of Physics and Astronomy \& Center for Physics and Chemistry of Living Systems,
Tel Aviv University, Ramat Aviv, Tel Aviv 69978, Israel}

\author{Shigeyuki Komura}\email{komura@wiucas.ac.cn\\
ORCID: 0000-0003-3422-5745}
\affiliation{Zhejiang Key Laboratory of Soft Matter Biomedical Materials, Wenzhou Institute, 
University of Chinese Academy of Sciences, Wenzhou, Zhejiang 325001, China}

\begin{abstract}
We review the theoretical development and modern applications of the Evans-Sackmann hydrodynamic model for lateral 
transport in supported fluid membranes.
We first cover the original formulation, emphasizing the linear momentum decay term that captures 
membrane-substrate coupling mediated by a thin lubricating fluid layer.
This coupling term enables quantitative interpretation of tracer diffusion measurements in supported bilayers.
We then survey theoretical extensions that relax standard boundary conditions 
at the inclusion perimeter.
Here, inclusions refer to embedded objects such as proteins, lipid domains, or tracer particles within the membrane.
We discuss the drag on a disk and on a liquid domain, as well as the dynamics of membrane phase separation. 
We also show that the supported-membrane mobility tensor provides a unified framework for correlated diffusion,
polymer dynamics, phase separation kinetics, and many-body interactions.
Finally, we discuss recent extensions to active and chiral membranes, where odd viscosity provides 
a transverse hydrodynamic response and offers a possible route for detecting chirality in two-dimensional fluids.
\end{abstract}


\maketitle


\section{Introduction}

Lipid bilayers are the main structural components of biomembranes and constitute flexible fluid-like interfaces in 
which many molecular transport processes occur effectively in two dimensions (2D)~\cite{singer1972fluid}.
Because bilayers are only a few nanometers thick, much smaller than their lateral dimensions,
the motion of lipids and membrane-associated objects is effectively confined to the membrane plane.
In the fluid-mosaic picture, lipids behave as a 2D in-plane fluid, allowing embedded or attached molecules to diffuse 
laterally along the membrane surface~\cite{singer1972fluid}.
Such membrane inclusions are, for example, proteins, lipid domains, and tracer particles, whose motion provides important 
information on the membrane hydrodynamic properties~\cite{albertsbook, edidin1976measurement, simons1997functional}.
Although modern views of biological membranes emphasize a complex, hierarchical organization beyond the fluid-mosaic model, this model still provides a useful framework for simplified membrane systems, including artificially reconstituted membranes, liposomes, and blebbed plasma membranes~\cite{kusumi2012dynamic}.

The inherent 2D character presents a significant theoretical challenge known as the Stokes' paradox~\cite{Landau1987, happel2012low}.
In a purely 2D incompressible viscous fluid, one cannot obtain a steady solution of the flow around a rigid body translating with 
a constant velocity when the fluid velocity satisfies a no-slip condition at the body perimeter and approaches zero at infinity.
This is because the velocity field cannot simultaneously satisfy the boundary conditions at the rigid object's surface and at infinity.
To resolve this paradox, one must introduce a mechanism for momentum decay from the 2D fluid.
In many experimental setups, a free-standing
membrane is surrounded by a bulk three-dimensional (3D) fluid solvent, which naturally 
provides this momentum decay~\cite{lipowsky1995structure}.
Consequently, the lateral mobility of such an inclusion is governed not only by its size and membrane viscosity, 
but also by the momentum exchange between the free-standing fluid membrane and the surrounding 3D solvent.

The theoretical framework for lateral transport in free-standing membranes was established by the pioneering 
hydrodynamic model of Saffman and Delbr\"{u}ck~\cite{saffman1975}.
Their main idea was to couple a 2D viscous fluid sheet with a surrounding 3D bulk fluid (e.g., water).
This resolves the Stokes' paradox and leads to a weak logarithmic dependence 
of the translational mobility on the inclusion size.
One of the central outcomes of this quasi-2D model is the existence of a hydrodynamic screening length known as 
the Saffman-Delbr\"{u}ck (SD) length.
This length scale separates the regime in which dissipation is dominated by 2D membrane viscosity from the regime 
in which the surrounding 3D fluid controls the flow~\cite{saffman1976}.
Subsequent theoretical works have extended the SD model to account for intermediate values of 
membrane viscosity~\cite{hughes1981translational}, membrane curvature~\cite{henle2010, shi2024drag}, 
and the correlated diffusion of inclusions such as membrane proteins~\cite{oppenheimer2009}.

\begin{figure}[htb]
\centering
\includegraphics[width=.75\linewidth]{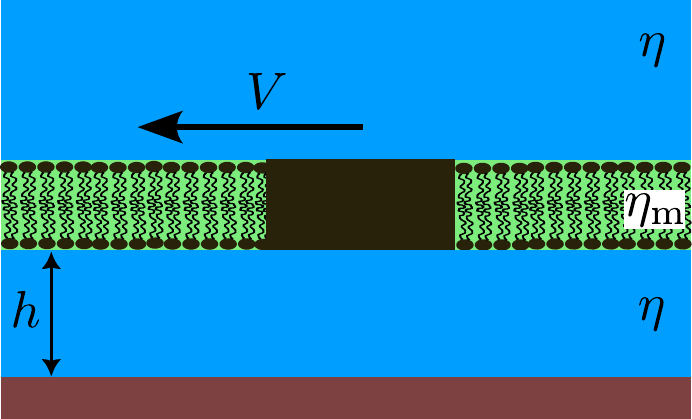}
\caption{
Schematic representation of a supported lipid bilayer membrane as in the Evans-Sackmann (ES) hydrodynamic model. 
The membrane is located between an upper semi-infinite bulk fluid and an underlying lubricating fluid 
layer of thickness $h$.
Both fluids are assumed to have the same 3D shear viscosity $\eta$, and the lubricating fluid velocity vanishes at the bottom substrate surface  (no-slip boundary condition). 
Due to the in-plane fluidity of the lipid membrane, an inclusion is moving with a constant velocity $V$ 
within the membrane plane.
This in-plane motion is modeled as transport in an effectively 2D fluid layer (thus, of vanishing thickness) with a 2D shear viscosity $\eta_{\rm m}$.}
\label{fig:membrane}
\end{figure}

\begin{figure*}[htb]
\centering
\includegraphics[width=.95\linewidth]{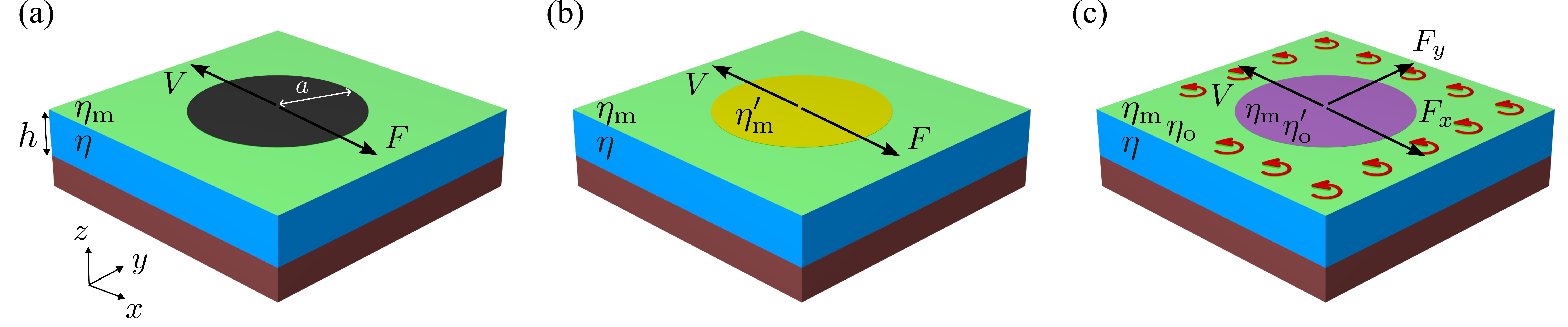}
\caption{
Schematic 3D representation of the ES hydrodynamic model and its extensions (see 
also Fig.~\ref{fig:membrane}).
(a) A circular disk-like object (black) of radius $a$ moves laterally with velocity $\mathbf{V}=(-V,0)$ 
in the plane of the supported membrane (see Sec.~\ref{sec:ESsub}). The disk experiences a drag force $F=\zeta V$ acting in the opposite direction, 
where $\zeta$ is the drag coefficient. The resulting drag coefficient is given in Eq.~\eqref{eq:ES}. 
(b) An extension of the ES model for a moving circular liquid domain (yellow) (see Sec.~\ref{sec:drop}).  
The liquid domain is characterized by a shear viscosity $\eta_{\rm m}^\prime$, which can differ from the membrane viscosity $\eta_{\rm m}$.
The resulting drag coefficient is given in Eq.~\eqref{frictionb}.
(c) An extension of the ES model for a moving circular liquid domain with an odd viscosity $\eta_{\rm o}^\prime$ (purple) in a 
supported membrane with an odd viscosity $\eta_{\rm o}$ (see Secs.~\ref{sec:oddsub1} and \ref{sec:oddsub2}).
Besides the translational drag force $F_x$, an orthogonal lift force $F_y$ arises when $\eta_{\rm o}\neq\eta_{\rm o}^\prime$.
The corresponding drag coefficient and lift coefficient are given in  Eqs.~\eqref{eq:drag} and \eqref{eq:lift}, 
respectively. 
}
\label{fig:system}
\end{figure*}

Although free-standing membranes are simpler to model, many widely used experimental setups are 
realized for membranes in the proximity of solid surfaces and boundaries.
Examples include supported lipid bilayers that lie very close to a solid surface~\cite{sackmann1996supported}, 
polymer-cushioned membranes~\cite{tanaka2005polymer,Castellana2006}, 
and membranes confined in microfluidic geometries~\cite{Ainla2013}.
Furthermore, cell membranes \textit{in vivo} are typically tethered to the underlying cytoskeleton or strongly adhered to neighboring cells~\cite{albertsbook}.
Hence, it is inappropriate to model such lipid membranes without considering their surroundings, because
in these bounded or confined environments, the fluid membrane undergoes frictional coupling with the adjacent lubricating layer and 
the rigid substrate.

To address this problem, Evans and Sackmann (ES) made a seminal contribution by introducing an effective friction 
between the membrane and a rigid substrate in the presence of a thin lubricating layer~\cite{evans1988}.
In their work, they included a linear drag that represents weak membrane-substrate coupling, 
and formulated an analytically tractable 2D hydrodynamic model for a supported membrane.
The recognition that momentum decay regularizes otherwise divergent 2D flow has had a major impact on studies 
of protein and domain diffusion~\cite{merkel1989molecular,sackmann1996supported}.
Their model is directly related to experiments conducted on supported membranes, 
which are often engineered to precisely tune the coupling to the solid surface~\cite{sackmann1996supported}.
The ES model is mathematically analogous to the Brinkman equation for flow in 3D porous media~\cite{Brinkman1949}. 
In contrast to the Brinkman equation, which is formulated for 3D systems, the ES model introduces a corresponding 
momentum-decay mechanism into a 2D fluid. 
This mechanism is essential for resolving Stokes' paradox, which arises specifically in 2D hydrodynamics.

Examples of such systems range from tightly supported membranes, where friction with the substrate is strong, to 
membranes separated from the substrate by polymer cushions or tethers, which increase the membrane-substrate distance and 
reduce friction~\cite{sackmann2000supported,tanaka2005polymer}.
Moreover, extensions of the ES approach have been widely employed to infer effective membrane viscosities 
from measured diffusion coefficients~\cite{stone1998hydrodynamics, camley2013diffusion}, as well as to account for interleaflet friction in supported bilayers~\cite{anthony2022systematic}.
Similar geometries have been studied in the context of the microrheology of fluid layers that are 
supported~\cite{komura2012lateral} or confined~\cite{komura2012anomalous, bar2017correlations} inside a viscoelastic medium.
The ES model has therefore become a standard framework for analyzing the diverse membrane systems described above.

This review discusses the ES model and its applications to lateral transport phenomena.
In Sec.~\ref{sec:ES}, we review the original ES model for a rigid disk-shaped object immersed 
in a supported membrane. 
We then introduce in Sec.~\ref{sec:phase} the steady motion of a circular liquid domain.
We also discuss how the solid support affects phase separation dynamics in multicomponent lipid membranes, 
where liquid-ordered and liquid-disordered domains form and subsequently grow through lateral diffusion and 
coalescence.
To highlight the versatility of the ES approach, we introduce in Sec.~\ref{sec:Mobility} the mobility tensor of a supported membrane. 
Specifically, we discuss its application to a broader range of systems, including 2D suspensions of passive particles, 2D polymer 
solutions, and active diffusion and propulsion phenomena within supported membranes.
Finally, in Sec.~\ref{sec:odd}, we briefly highlight the recent theoretical development regarding the flow properties 
of 2D chiral active fluids characterized by a dissipationless transport coefficient called odd viscosity.

\section{Evans-Sackmann model for a supported membrane}
\label{sec:ES}

\subsection{Hydrodynamic equations}

We first review the supported membrane model proposed by Evans and 
Sackmann (ES) in 1988~\cite{evans1988}.
They formulated a minimal hydrodynamic theory for the lateral motion of a rigid body embedded in a fluid membrane. 
As shown in Fig.~\ref{fig:membrane}, the membrane is dynamically coupled to an underlying rigid substrate through an 
intervening lubricating layer.
The key novelty of this approach lies in replacing detailed near-wall 3D hydrodynamics with a simple momentum decay mechanism.
This momentum decay circumvents the Stokes' paradox, making the problem of the 2D body motion analytically tractable.

In the ES model, the lipid bilayer is viewed as an infinitely thin and flat incompressible 
2D Newtonian fluid with a shear viscosity $\eta_{\rm m}$, whose physical units are N$\cdot$s/m.
The membrane is characterized by an in-plane velocity field $\mathbf{v}(\mathbf{r})$ and pressure 
$p(\mathbf{r})$, where $\mathbf{r}=(x,y)$ denotes the 2D position vector with distance $r=|\mathbf{r}|$.
A thin lubricating fluid layer of thickness $h$ and 3D shear viscosity $\eta$ resides between the membrane and the substrate,
where $\eta$ has the usual physical units of N$\cdot$s/m$^2$. 
The ratio $\ell=\eta_{\rm m}/\eta$ therefore has the dimension of length and defines the Saffman-Delbr\"{u}ck (SD) length.
A no-slip boundary condition is imposed at the bottom substrate surface, 
and we assume that the lubricating layer is sufficiently thin such that $h \ll \ell$.

At low Reynolds numbers and in the steady state, the governing equations take the form of the in-plane force balance: 
\begin{align}
	\eta_{\rm m} \nabla^2\mathbf{v} - \nabla p - \frac{\eta}{h} \mathbf{v} = \mathbf{0},
	\label{eq:ES_stokes}
\end{align}
where $\eta/h$ is the membrane-substrate friction per unit area and accounts for momentum decay.
Physically, the term $(\eta/h) \mathbf{v}$ represents the areal force density required to shear 
a lubricating layer to overcome interfacial friction between the membrane and substrate.

This specific form of friction corresponds to dynamic coupling with an underlying substrate. 
More broadly, however, the momentum-decay term can represent various frictional interactions, such as 
those between lipid head groups and the surrounding environment, giving it wider physical significance.
Although there also exists a momentum decay into the semi-infinite bathing fluid above the membrane (see Fig.~\ref{fig:membrane}), 
it is negligible compared to that of the lubricating layer between the membrane and substrate~\cite{evans1988}.
Hence, the term $(\eta/h) \mathbf{v}$ with small $h$ dominates in Eq.~\eqref{eq:ES_stokes}.
As mentioned before, Eq.~\eqref{eq:ES_stokes} is mathematically analogous to the Brinkman equation 
for 3D porous media, where the momentum-decay term describes the frictional screening of hydrodynamic 
interactions~\cite{Brinkman1949}.

The hydrodynamic equation~\eqref{eq:ES_stokes} is supplemented by the 
incompressibility condition:
\begin{align}
\nabla \cdot \mathbf{v}=0.
\label{eq:incomp}
\end{align}
This equation arises because membranes behave as condensed liquid crystals with negligible area compressibility~\cite{evans1988}.
The key parameter in the ES model is the ES screening length, defined as
\begin{align}
\kappa^{-1} = \sqrt{ \frac{\eta_{\rm m}h}{\eta} }= \sqrt{h \ell}.
\label{screeninglength}
\end{align}
In contrast to the SD length $\ell$, the ES screening length $\kappa^{-1}$ depends on the thickness 
$h$ of the lubricating layer and directly reflects the strength of membrane-substrate coupling.
Hereafter, we use $\kappa$ to nondimensionalize all lengths.

At short distances $(\kappa r\ll 1)$, the friction term is weak, and the flow field resembles that 
of a 2D viscous sheet with conserved momentum.
Conversely, at large distances $(\kappa r\gg 1)$, momentum decays efficiently, and in-plane flows are strongly attenuated.
This crossover behavior constitutes the central physics of the ES model.
Due to the friction term, the long-range 2D hydrodynamics is regularized with a finite screening length
$\kappa^{-1}$, beyond which momentum decays into the surrounding environment.

\begin{figure}
\centering
\includegraphics[width=.75\linewidth]{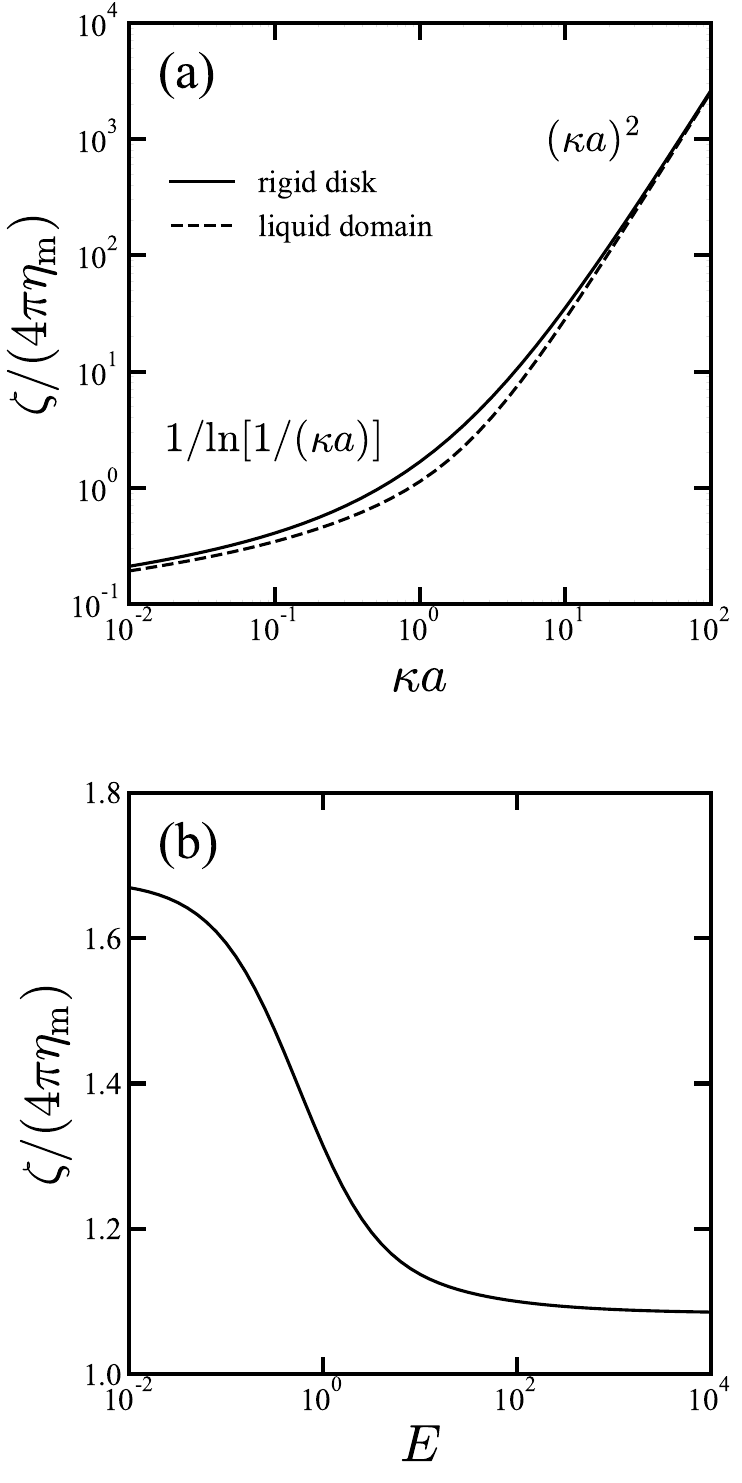}
\caption{
(a) Log-log plot of the dimensionless drag coefficient $\zeta/(4\pi \eta_{\rm m})$ 
as a function of the dimensionless size $\kappa a$ for a rigid disk [solid line, see Eq.~\eqref{eq:ES}] and 
for a liquid domain [dashed line, see Eq.~\eqref{frictionb}]. 
For the latter, the viscosity ratio is chosen as $E=\eta_{\rm m}/\eta_{\rm m}^\prime =10$.
The drag coefficient $\zeta$ shows a logarithmic dependence on size when $\kappa a \ll 1$, whereas  
it increases quadratically when $\kappa a \gg 1$.
(b) Semi-log plot of the dimensionless drag coefficient $\zeta/(4\pi \eta_{\rm m})$ for a liquid domain 
as a function of the viscosity ratio $E$.
The liquid domain size is chosen as $\kappa a =1$. 
}
\label{fig:drag}
\end{figure}

\subsection{Drag coefficient for a disk}
\label{sec:ESsub}

Evans and Sackmann analyzed the steady translation of a rigid disk of radius $a$ moving with constant velocity 
$\mathbf{V}=(-V,0)$ in a supported membrane, as shown in Fig.~\ref{fig:system}(a).
For the boundary conditions, they imposed a no-slip condition on the disk perimeter 
and assumed that the flow velocity decays to zero at infinity.
In the presence of the ES screening length $\kappa^{-1}$, the flow field can be expressed explicitly 
in terms of modified Bessel functions of the second kind of order $n$, $K_n(z)$~\cite{abramowitz2000handbook}.
Considering the drag force $F$ and drag coefficient $\zeta$ through the linear relation $F=\zeta V$, they obtained the 
dimensionless drag coefficient~\cite{evans1988}
\begin{equation}
	\frac{\zeta}{4\pi\eta_{\rm m}}=
	\frac{(\kappa a)^2}{4} + 
	\frac{\kappa a K_1(\kappa a)}{K_0(\kappa a)}.
	\label{eq:ES}
\end{equation}
See Appendix~\ref{app} for the explicit derivation of $\zeta$, and 
Fig.~\ref{fig:drag}(a) for the plot of $\zeta$ as a function of the disk size.

The analytical expression in Eq.~\eqref{eq:ES} admits two limiting regimes.
For small disks or weak screening ($\kappa a \ll 1$), the drag coefficient exhibits a weak logarithmic dependence on the disk size:
\begin{align}
	\frac{\zeta}{4\pi\eta_{\rm m}} \approx \frac{1}{\ln[2/(\kappa a)]-\gamma},
	\label{logdependent}
\end{align}
where $\gamma= 0.5772...$ denotes Euler's constant.
In this limit, the first quadratic term in Eq.~(\ref{eq:ES}) is negligible since it vanishes algebraically as $\kappa a \to 0$, 
whereas the logarithmic term varies much more slowly.
Such a logarithmic behavior is characteristic of purely 2D hydrodynamics~\cite{happel2012low} 
and is similar to the translational drag derived for a free-standing membrane~\cite{saffman1975}.

For large disks or strong screening ($\kappa a \gg 1$), on the other hand, the response in Eq.~\eqref{eq:ES} 
becomes friction-dominated,  and the drag coefficient shows a quadratic dependence on size:
\begin{align}
	\frac{\zeta}{4\pi\eta_{\rm m}} \approx \frac{(\kappa a)^2}{4}.
	\label{eq:ESlrg}
\end{align}
This quadratic dependence contrasts with the free-standing membrane case, where the drag coefficient 
scales linearly with the size $a$ due to the 3D hydrodynamics of the surrounding bulk fluid~\cite{diamant2009hydrodynamic}.

The drag coefficient is linked to the diffusion coefficient of the circular disk via Einstein's relation, 
$D = k_{\rm B}T/\zeta$, where $D$ is the diffusion coefficient, $k_{\rm B}$ is the Boltzmann constant, 
and $T$ is the temperature~\cite{doi2013soft}.
Analyzing the size dependence of the diffusion coefficient, therefore, 
provides a powerful experimental tool for probing the properties of supported membranes. 
Specifically, for large diffusing objects ($\kappa a \gg 1$), such as large lipid domains, colloidal particles, or membrane-bound beads,
the quadratic dependence prevails in the translational drag, as in Eq.~\eqref{eq:ESlrg}.
In this limit, the diffusion coefficient scales as $D\approx k_{\rm B}T h/(\pi\eta a^2)$, 
and does not depend on the membrane viscosity $\eta_{\rm m}$.
This expression allows several physical quantities to be extracted experimentally, such as 
the disk size $a$ or the lubrication-layer thickness $h$~\cite{sackmann1996supported}.

The crossover from the logarithmic to algebraic behaviors of the drag coefficient takes place when $a \approx \kappa^{-1}$.   
A typical length scale corresponding to this condition can be roughly estimated. 
From Eq.~(\ref{screeninglength}), we have $\kappa^{-1} \approx 10^{-7}$\,m $=0.1$\,$\mu$m for 
$\eta_{\rm m} \approx 10^{-9}$\,N$\cdot$s/m, 
$\eta \approx 10^{-3}$\,N$\cdot$s/m$^2$, and
$h=10^{-8}$\,m.
Such a crossover length scale $a \approx 0.1$\,$\mu$m is accessible in many experiments.

\section{Liquid domains in a supported membrane}
\label{sec:phase}

\subsection{Drag coefficient for a liquid domain}
\label{sec:drop}

Because the Evans-Sackmann (ES) hydrodynamic model provides a versatile yet mathematically 
simple description of supported membranes, numerous theoretical extensions have been developed.
These include generalizations to inter-bilayer friction~\cite{seki2014diffusion, hill2014diffusion}, weakly compressible
membranes~\cite{barentin1999, elfring2016surface, manikantan2020surfactant, hosaka2021nonreciprocal, 
hosaka2023pair, hosaka2023hydrodynamics, lier2023lift, daddi2025analytical, daddi2025hydrodynamic}, 
spherical geometries~\cite{shi2024drag}, arbitrary-thickness sublayers~\cite{stone1998hydrodynamics, seki2011diffusion}, 
and Brownian dynamics~\cite{seki1993brownian}.

We now discuss the role of boundary conditions at the inclusion perimeter.
Most previous studies using the ES framework have examined the viscous resistance of rigid objects subject to no-slip boundary conditions.
However, this assumption is not always applicable to all membrane inclusions.
For instance, lipid molecules are often distributed non-uniformly within the membrane to form lipid-rich functional 
domains~\cite{simons1997functional, veatch2005seeing, komura2012dynamics, komura2014physical}.
In such liquid domains, called lipid rafts~\cite{simons1997functional,Lingwood2010}, the internal viscosity can differ from that of the surrounding membrane, 
giving rise to a finite interfacial surface velocity at the domain boundary.

To elucidate the influence of boundary conditions, Ramachandran \textit{et al}.\ derived the drag coefficient 
of a circular liquid domain characterized by a shear viscosity, $\eta_{\rm m}^\prime$ 
[see Fig.~\ref{fig:system}(b)]~\cite{ramachandran2010drag}.
Unlike a rigid inclusion, a liquid domain can sustain internal circulation, and this internal flow modifies the interfacial stress balance 
at the domain boundary.
The translational resistance in this situation exhibits complex behavior, depending not only on the domain size $a$, 
but also on the viscosity ratio, $E= \eta_{\rm m}/\eta_{\rm m}^\prime$.
By neglecting domain deformations, the analytical expression for the scaled drag coefficient is given by~\cite{ramachandran2010drag}
\begin{align}
\frac{\zeta}{4\pi \eta_{\rm m}} = 
\frac{(\kappa a)^2}{4} +
\frac{ \kappa a  ( \kappa' a I'_1 - 2 I'_2+ 2 E I'_2)K_1}
{(\kappa' a I'_1 - 2 I'_2)K_0 + E(2 K_0+\kappa a K_1)I'_2},
\label{frictionb}
\end{align}
where $\kappa'  = \sqrt{E} \kappa $. For brevity, the arguments of the modified Bessel functions are omitted, 
such that $K_n = K_n(\kappa a)$ and $I_n' = I_n(\kappa' a)$.
The asymptotic expression of Eq.~\eqref{frictionb} in the limit of $E \to 0$ (larger  
$\eta_{\rm m}^\prime$ of the domain) recovers the rigid disk behavior, Eq.~\eqref{eq:ES}, as it should.
The opposite limit of $E\to\infty$ (smaller $\eta_{\rm m}^\prime$) 
corresponds to the case of a 2D gas bubble domain, and its drag coefficient is given by~\cite{ramachandran2010drag}
\begin{align}
	\frac{\zeta}{4\pi \eta_{\rm m}} \approx
	\frac{(\kappa a)^2}{4}
	+
	\frac{2 \kappa a K_1}{2K_0+\kappa a K_1}.
\end{align}

As shown in Fig.~\ref{fig:drag}(a) when $E=10$, the drag coefficient of a liquid domain is smaller compared 
to that of the rigid disk. 
Moreover, as plotted in Fig.~\ref{fig:drag}(b), the drag decreases monotonically with increasing $E$. 
This reduction is attributed to the internal flows generated in the liquid domain, 
which facilitate more efficient fluid transport~\cite{ramachandran2010drag}.

Multicomponent lipid membranes can laterally demix into coexisting liquid phases~\cite{Dietrich2001}.
A typical example is the separation between a liquid-ordered (Lo) phase, which is enriched in saturated 
lipids and cholesterol, and a liquid-disordered (Ld) phase, which is enriched in unsaturated lipids.
This Lo-Ld phase separation provides a useful model for lipid rafts~\cite{simons1997functional,Lingwood2010}, 
and has been widely studied in giant vesicles and supported membranes~\cite{Veatch2003,veatch2005seeing}.
In such phase-separated membranes, the domain viscosity can differ from that of the surrounding membrane, and this viscosity 
contrast affects the hydrodynamic drag of a liquid domain~\cite{Cicuta2007}.
By investigating the phase separation between Lo and Ld phases, 
realistic values for the ratio $E$ were estimated to be $0.1 \lesssim E \lesssim 10$~\cite{Oradd2005}.

\subsection{Growth law of phase-separated domains}

In addition to equilibrium phase behavior, the coarsening dynamics of phase-separated domains 
have also been investigated in multicomponent membranes~\cite{Baumgart2003,Yanagisawa2007,Stanich2013}.
After a quench into the two-phase region, small domains nucleate and subsequently grow in order to reduce the total line 
energy associated with the Lo-Ld domain boundaries~\cite{Dietrich2001,Veatch2003,veatch2005seeing}.
Because both Lo and Ld phases are liquid, these domains can diffuse laterally, collide, and coalesce within the membrane plane.
The resulting coarsening dynamics are therefore governed not only by line tension but also by the size-dependent domain mobility.
In supported membranes, this mobility is strongly affected by hydrodynamic screening due to the nearby substrate, and it can 
affect domain growth dynamics.

According to the dynamic scaling hypothesis, there exists a regime where the average domain size 
$R$ grows with time $t$ as $R\sim t^\alpha$, where $\alpha$ is a universal growth exponent~\cite{Bray2002,Onuki2002}.
Several distinct mechanisms can contribute to domain coarsening, including Brownian coagulation, Ostwald ripening, and 
hydrodynamic flow-induced coalescence.
Here, we focus on Brownian coagulation, which is directly controlled by domain mobility.
For the 3D Brownian coagulation process, this exponent is known to be $\alpha=1/3$~\cite{Bray2002,Onuki2002}.

The hydrodynamic screening that modifies the mobility of an isolated domain affects collective coarsening, 
because Brownian coagulation is controlled by the size dependence of domain diffusion.
Using dissipative particle dynamics simulations, Ramachandran \textit{et al.}\
examined how hydrodynamic coupling to the surrounding bulk fluid modifies the phase separation kinetics 
in a thin planar liquid membrane~\cite{ramachandran2010effects}.
For off-critical mixtures in purely 2D systems, their simulations showed a growth exponent of $\alpha=1/2$, 
different from the 3D exponent of $1/3$.

This discrepancy can be understood through the Brownian coagulation mechanism.
Dimensional analysis for a 2D domain yields a diffusion coefficient $D\sim k_{\rm B}T/\eta_{\rm m}$, which  
is independent of the domain size or depends only logarithmically on it, 
as shown in Eq.~\eqref{logdependent}.
Combining this with the scaling relation $R^2\sim Dt\sim(k_{\rm B}T/\eta_{\rm m})t$, 
one obtains $R\sim t^{1/2}$, which is consistent with the simulation results.
This argument applies to both supported and free-standing membranes as long as the relevant domain size is 
smaller than the hydrodynamic screening length (either SD or ES screening length).

Applying a similar scaling argument, they further explored the influence of a nearby substrate 
when the domain size becomes large.
Since $D\sim R^{-2}$ at large length scales in a supported membrane,
as shown in Eq.~\eqref{eq:ESlrg}, they predicted that the domain growth exponent becomes $\alpha=1/4$
in the presence of a solid substrate.
In contrast, when hydrodynamic interactions are governed by the surrounding 3D bulk fluid, as in free-standing 
membranes, the domain diffusion coefficient scales as $D \sim R^{-1}$ and the coarsening exponent becomes $\alpha = 1/3$,
recovering the classical 3D value even in 2D free-standing membranes.

\begin{figure*}[tbh]
\begin{center}
\includegraphics[width=.95\linewidth]{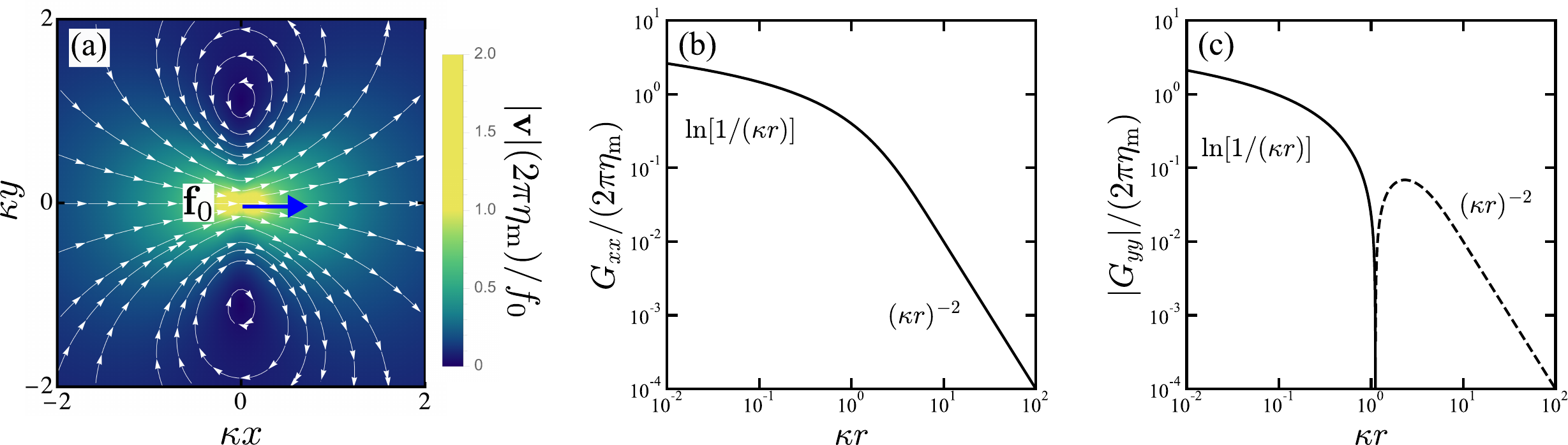}
\end{center}
\caption{
(a) Streamlines of the fluid velocity field in the $(x,y)$-plane of a 2D supported membrane. 
The velocity field induced by a force monopole $\mathbf{f}_0=f_0\hat{\mathbf {x}}$ at the origin 
is described by the mobility tensor in Eq.~\eqref{eq:mobility_es}.
The color indicates the magnitude of the rescaled velocity, $\vert \mathbf v \vert (2\pi\eta_{\rm m})/f_0$.
Notice that the vortex centers are located at $\kappa y \approx \pm 1$.
(b) Log-log plot of the longitudinal mobility component $G_{xx}$ (rescaled by $2\pi\eta_{\rm m}$) as a function of $\kappa r$.
(c) Log-log plot of the transverse mobility component $G_{yy}$ as a function of $\kappa r$.
Since $G_{yy}$ becomes negative for $\kappa r \gtrsim 1$ in the log-log plot, we have plotted its 
absolute value as shown by the dashed line.
Both $G_{xx}$ and $G_{yy}$ show a logarithmic dependence for $\kappa r \ll 1$, whereas  
they decay quadratically for $\kappa r \gg 1$. 
\label{fig:G}
}
\end{figure*}

\subsection{Dynamics of critical concentration fluctuations}

Besides domain coarsening below the miscibility transition temperature, 
concentration fluctuations above the transition temperature have also attracted considerable attention in multicomponent lipid membranes~\cite{Veatch2007Critical,HonerkampSmith2008,Veatch2008GPMV}.
Near the critical point, transient composition fluctuations emerge over a broad range of length and time scales, 
and their dynamics are strongly influenced by membrane hydrodynamics.
Experimental studies on ternary lipid mixtures have demonstrated that these critical fluctuations exhibit behavior consistent with the 
2D Ising universality class~\cite{HonerkampSmith2008}.

The dynamics of concentration fluctuations in multicomponent membranes were theoretically investigated by 
Seki \textit{et al.}\ using the ES hydrodynamic model~\cite{seki2007concentration}.
Within this framework, the decay rate of concentration fluctuations with wavenumber $q$ is given by 
$D_q q^2$, where $D_q$ is the wavenumber-dependent diffusion coefficient, and the mode relaxes as $\exp(-D_q q^2 t)$.
The asymptotic forms of $D_q$ were obtained as  
$D_q \approx k_{\rm B}T/(4\pi\eta_{\rm m})\ln(q/\kappa)$ for $q/\kappa \gg 1$, 
and $D_q \approx k_{\rm B}T/(4\pi\eta_{\rm m})(q/\kappa)^2$ for $q/\kappa \ll 1$.
These two asymptotic forms are also consistent with the size dependence of the diffusion coefficient 
of a membrane inclusion if its characteristic size is identified as $a \sim 1/q$.
Short-wavelength fluctuations ($q/\kappa \gg 1$) correspond to the weakly screened regime, in which hydrodynamic interactions are 
effectively 2D, whereas long-wavelength fluctuations ($q/\kappa \ll 1$) correspond to the friction-dominated regime.

Concentration fluctuations in multicomponent membranes were further investigated in experimentally more relevant geometries, 
such as supported membranes with finite-thickness solvents~\cite {ramachandran2011hydrodynamic}.
An interesting extension of this framework concerns 2D microemulsions in lipid membranes, motivated by the lineactant effect of hybrid lipids.
In this case, the static structure factor develops a peak at a finite wavenumber, leading to characteristic non-monotonic behavior 
in the effective diffusion coefficient $D_q$.

\section{Hydrodynamic response in a supported membrane}
\label{sec:Mobility}

\subsection{Mobility tensor of a supported membrane}
\label{sec:Mobilitysub}

The Evans-Sackmann (ES) model is particularly useful because it provides an explicit mobility tensor for membranes 
with screened hydrodynamic interactions. 
This tensor relates a local in-plane force to the resulting membrane flow and describes how membrane inclusions interact with each other hydrodynamically through the supported membrane.
More generally, the mobility tensor provides the basis for analyzing transport phenomena, far-field flows generated by passive 
and active inclusions~\cite{lauga2020fluid}, long-range hydrodynamic interactions~\cite{happel2012low}, and numerical methods 
such as the boundary element method~\cite{pozrikidis1992}. 
Here, active inclusions refer to macromolecules, such as enzymes or proteins, that continuously consume energy to generate motion or mechanical stresses. In this section, we discuss three representative applications of the mobility tensor of a supported membrane: the lateral dynamics of passive inclusions (including correlated diffusion and 2D polymer dynamics), the lateral dynamics induced by active inclusions with force dipoles, and the locomotion of a 2D micro-swimmer.

Solving the ES hydrodynamic equation~\eqref{eq:ES_stokes} for $\mathbf{v}$ in Fourier space 
and performing the inverse Fourier transform yields the real-space representation of the mobility tensor 
(or Green's function) $\mathbf{G}(\mathbf{r})$ for a supported membrane.
For a general in-plane force density $\mathbf f(\mathbf r)$, the induced membrane velocity is given by the  
convolution $\mathbf v(\mathbf r)=\int {\rm d}\mathbf r'\,\mathbf G(\mathbf r-\mathbf r')\cdot \mathbf f(\mathbf r')$,
where $\mathbf{G}(\mathbf{r})$ is obtained as~\cite{ramachandran2011, komura2012dynamics}
\begin{align}
	\mathbf{G}(\mathbf{r})
	&=\frac{1}{2\pi\eta_{\rm m}}\left[
	K_0(\kappa r)+\frac{K_1(\kappa r)}{\kappa r}-\frac{1}{(\kappa r)^2}
	\right]\mathbf{I}
	\notag\\
	&+\frac{1}{2\pi\eta_{\rm m}}\left[
	-K_0(\kappa r)-\frac{2K_1(\kappa r)}{\kappa r}+\frac{2}{(\kappa r)^2}
	\right]\frac{\mathbf{r\,r}}{r^2},
	\label{eq:mobility_es}
\end{align}
with $\mathbf{I}$ being the 2D identity tensor. 
A full derivation of this mobility tensor is provided in Appendix~\ref{appb}.
The velocity field induced by a force monopole at the origin $\mathbf{f}(\mathbf{r})=\mathbf{f}_0\delta(\mathbf{r})$
is depicted in Fig.~\ref{fig:G}(a).

By expanding $\mathbf{G}(\mathbf{r})$ for small $r$, we obtain the limiting expression for $\kappa r\ll1$~\cite{oppenheimer2010}
\begin{align}
	\mathbf{G}(\mathbf{r})
	\approx
	\frac{1}{4\pi\eta_{\rm m}}\left[
	\left(
	\ln\frac{2}{\kappa r} - \gamma -  \frac{1}{2}
	\right)\mathbf{I}
	+
	\frac{\mathbf{r\,r}}{r^2}
	\right],
	\label{eq:mobility_es_sml}
\end{align}
whereas for $\kappa r\gg1$, the asymptotic expression becomes 
\begin{align}
	\mathbf{G}(\mathbf{r})
	\approx
	-\frac{1}{2\pi\eta_{\rm m}(\kappa r)^2}
	\left(\mathbf{I}-2\frac{\mathbf{r\,r}}{r^2}\right).
	\label{eq:mobility_es_lrg}
\end{align}
The mobility tensor exhibits a logarithmic dependence at short distances and decays algebraically as $r^{-2}$ at 
large distances.

For a point force applied along the $x$-direction, as shown in Fig.~\ref{fig:G}(a), the velocity component along the force is governed by the longitudinal $G_{xx}$ and transverse $G_{yy}$ components, which are evaluated for separation vectors $\mathbf{r}$ parallel or perpendicular to the direction of the force, respectively.
Figures~\ref{fig:G}(b) and (c) show their full behavior as a function of $\kappa r$.
In contrast to the monotonically decaying longitudinal component $G_{xx}$, the transverse component $G_{yy}$ becomes negative around $\kappa r \approx 1.11$ and vanishes at infinity.
As shown in Fig.~\ref{fig:G}(a), the sign reversal of $G_{yy}$ reflects the recirculating flow induced by momentum screening 
in a quasi-2D incompressible fluid.

\begin{figure*}[tbh]
\begin{center}
\includegraphics[width=0.95\linewidth]{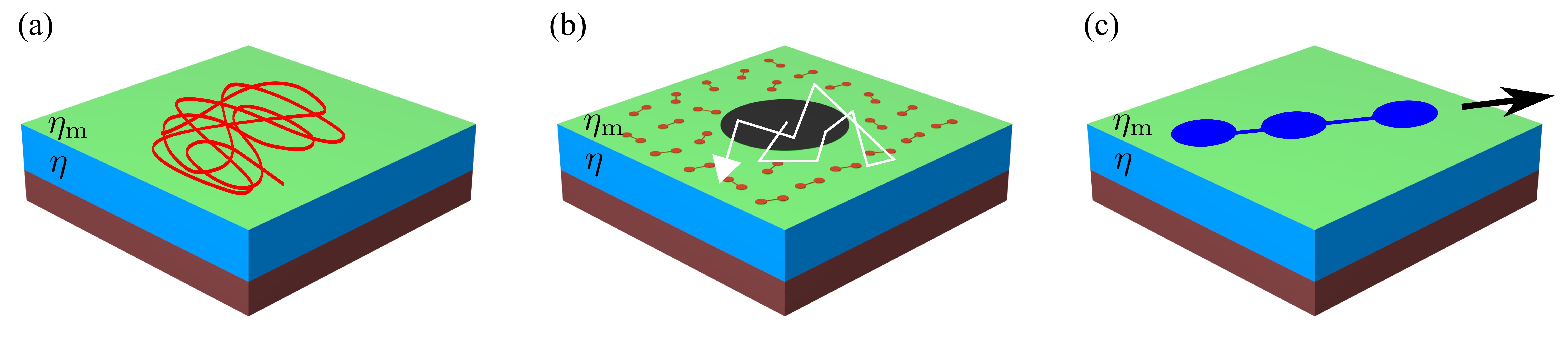}
\end{center}
\caption{
(a) A polymer chain (red) embedded in a supported membrane (see Sec.~\ref{sec:passive}).
(b) A passive inclusion (black) diffusing laterally in a supported membrane that contains active inclusions 
such as enzymes (see Sec.~\ref{sec:active}). 
Active inclusions embedded in the membrane are represented as active force dipoles (red dimers). 
The passive inclusion undergoes Brownian motion due to thermal and non-thermal fluctuations that 
are induced by active force dipoles.
(c) A three-disk micro-swimmer (blue) moving laterally in a supported membrane (see Sec.~\ref{sec:swimmer}).
The swimmer consists of three disks that are connected by two arms of variable lengths. 
\label{fig:dipole}
}
\end{figure*}

\subsection{Lateral dynamics of passive inclusions}
\label{sec:passive}

Using the exact mobility tensor of Eq.~\eqref{eq:mobility_es}, several studies have explored lateral hydrodynamics in 
supported membranes~\cite{komura1995diffusion, ramachandran2011}.
For instance, Oppenheimer \textit{et al.}\ derived the coupling diffusion coefficients 
for membrane inclusions in the far-field limit~\cite{oppenheimer2010}.
They analyzed how inclusions modify the hydrodynamic response and calculated 
the leading-order concentration corrections to the coupling diffusion coefficients.

Komura \textit{et al.}\ obtained analytical expressions 
for the diffusion coefficient of a 2D polymer chain immersed in a membrane within the ES hydrodynamic model
[see Fig.~\ref{fig:dipole}(a)]~\cite{komura1995diffusion}.
This was later extended by Ramachandran \textit{et al.}, who developed a Brownian dynamics theory incorporating full 
hydrodynamics for a 2D Gaussian polymer chain embedded in a membrane confined by bulk solvent and rigid walls~\cite{ramachandran2011}. 
For large polymer chain length, the relaxation time 
and the dynamical structure factor show Rouse-like behavior for membranes in confined geometries and
for supported membranes.
In contrast, the behavior reduces to Zimm-like behavior in free-standing membranes.
Furthermore, scaling arguments were employed in order to elucidate the effect of excluded-volume interactions on 
the 2D polymer relaxation time.

\subsection{Lateral dynamics induced by active inclusions}
\label{sec:active}

The mobility tensor, originally used to analyze the dynamics of passive inclusions, can be  
used to model lateral transport induced by active inclusions in membranes.
These active inclusions span a wide range of scales, from enzymatic proteins on the nanometer scale~\cite{mikhailov2015} 
to micro-organisms and synthetic micro-swimmers at the micrometer scale~\cite{lauga2020fluid}.
During catalytic cycles, enzymes undergo cyclic conformational changes driven by substrate binding and product release. 
In a fluidic environment, these nonequilibrium processes generate long-range dipolar flow fields, 
leading to non-thermal collective effects in the cytoplasm and in other sub-cellular organelles~\cite{mikhailov2015}.

To leading order, the hydrodynamic disturbance generated by an enzyme is represented as a force dipole, namely, 
a pair of equal and opposite point force monopoles.
Within the far-field approximation, the flow due to a single force dipole is calculated 
by taking the spatial derivative of the mobility tensor~\cite{lauga2020fluid}. 
Mikhailov \textit{et al.}\ analyzed the advective transport of passive molecules induced by active dipolar proteins 
in free-standing membranes~\cite{mikhailov2015}.
Using a mobility tensor with a distance dependence similar to Eq.~\eqref{eq:mobility_es_sml}, 
they derived the diffusion coefficient of a small passive probe in a free-standing membrane.
The effect is inherently nonlocal, meaning that active inclusions distributed throughout 
the membrane will contribute to the local enhancement of diffusion~\cite{mikhailov2015}.

More recently, Hosaka \textit{et al.}\ extended these results by considering a supported membrane [see Fig.~\ref{fig:dipole}(b)], 
and derived the active diffusion coefficient for inclusions of size $a$~\cite{hosaka2017}.
For a small inclusion, the active diffusion coefficient exhibits a logarithmic dependence on size, as for the passive inclusions 
discussed above.
In the opposite limit of a large inclusion size, however, the active diffusion coefficient decays algebraically as $a^{-4}$. 
This decay is much faster than that of the thermal diffusion coefficient, which decays as $a^{-2}$.
Manikantan further showed that substrate-induced hydrodynamic screening can tune the collective organization of active 
inclusions in viscous membranes~\cite{manikantan2020}.

\subsection{2D micro-swimmers in a supported membrane}
\label{sec:swimmer}

According to the scallop theorem for micro-swimmers~\cite{purcell2014life}, a force dipole undergoing reciprocal motion 
cannot achieve net propulsion in a viscous fluid.
However, internal non-reciprocal conformational changes, 
in which the sequence of shapes differs between forward and backward time evolution, 
can break this symmetry, generating net motion.

Ota \textit{et al.}\ proposed a three-disk micro-swimmer model in which three collinear rigid disks, 
connected by two arms, swim in the plane of a 2D supported membrane [see Fig.~\ref{fig:dipole}(c)]~\cite{ota2018three}.
In this model, the arm lengths vary in time according to a prescribed non-reciprocal actuation cycle, which provides the 
internal driving required for propulsion.
Taking into account the self-mobility of a rigid disk, Eq.~\eqref{eq:ES}, and the inter-disk hydrodynamic interaction, 
Eq.~\eqref{eq:mobility_es}, they showed that the average swimming speed exhibits three different asymptotic behaviors~\cite{ota2018three}.
Their analysis further revealed that the maximum velocity is obtained when the disks are equal in size, whereas it is minimized 
when the average arm lengths are identical.
Similar qualitative trends are also observed for a three-sphere micro-swimmer immersed in a 3D fluid when the corresponding 
geometries are the same~\cite{najafi2004,golestanian2008}. 
However, the dependence on the disk-size ratio is much weaker for 2D swimmers~\cite{ota2018three}. 
This weak dependence arises from the characteristic logarithmic nature of 2D hydrodynamics.

\section{A supported membrane with broken symmetries}
\label{sec:odd}

\begin{figure*}[htb]
\centering
\includegraphics[scale=0.36]{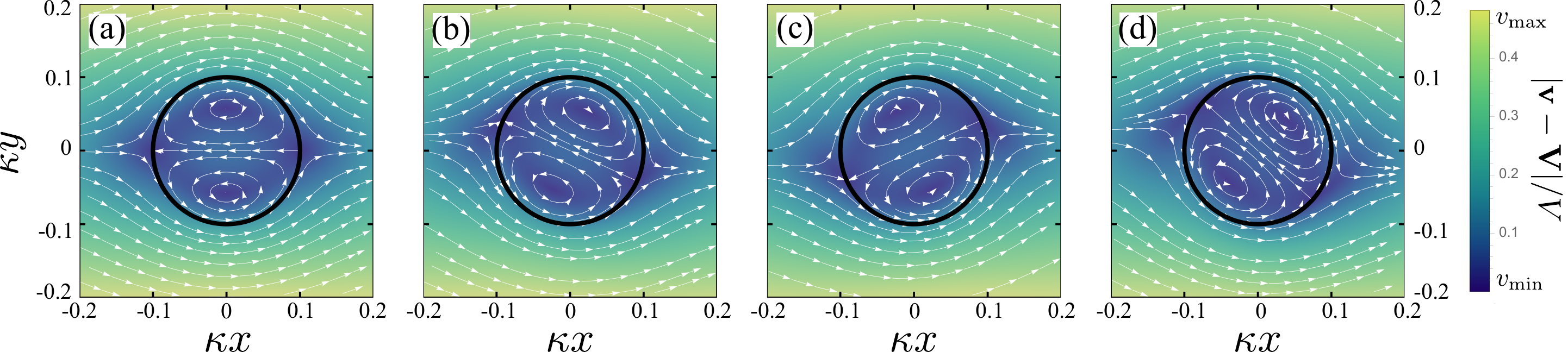}
\caption{
Translational motion of a circular liquid domain with an odd viscosity $\eta_{\rm o}^\prime$ in a 
supported membrane with an odd viscosity $\eta_{\rm o}$, as represented in Fig.~\ref{fig:system}(c).
The domain moves laterally in the negative $x$-direction with a velocity $\mathbf{V}=(-V,0)$.
Streamlines of the fluid velocity are shown in the comoving frame, $\mathbf{v}-\mathbf{V}$, 
when (a) $\eta_{\rm o}=\eta_{\rm o}^\prime=\eta_{\rm m}$ (uniform odd viscosity), 
(b) $\eta_{\rm o}=\eta_{\rm m}$ and $\eta_{\rm o}^\prime=0$ (vanishing odd viscosity inside the domain), 
(c) $\eta_{\rm o}=0$ and $\eta_{\rm o}^\prime=\eta_{\rm m}$ (vanishing odd viscosity outside the domain), 
and (d) $\eta_{\rm o}=-\eta_{\rm o}^\prime=\eta_{\rm m}$ (positive odd viscosity outside the domain and negative odd viscosity inside).
We have chosen the same shear viscosity inside and outside the domain, $E=\eta_{\rm m}/\eta_{\rm m}^\prime=1$, 
and $\kappa a=0.1$.
The circular black line represents the perimeter of the liquid domain. 
The color indicates the magnitude of the rescaled velocity with the minimum 
and maximum velocities $v_{\rm min}=0$ and  $v_{\rm max}=0.5$, respectively. 
}
\label{fig:3}
\end{figure*}

\subsection{2D odd viscosity}

Although active objects in fluid membranes are often modeled as single or multiple 2D particles translating through a Newtonian fluid, 
collective effects can fundamentally change the macroscopic flow properties.
For instance, in 2D suspensions of rigid disks, the effective membrane viscosity increases with disk concentration~\cite{oppenheimer2009}.
In contrast to these passive systems, the influence of active inclusions on membrane properties remains largely unexplored.
Key active constituents in this context include ATP synthase or ${\rm F}_0{\rm F}_1$-ATPase, 
which exhibit intrinsic rotational motion within biological membranes~\cite{albertsbook, hosaka2022nonequilibrium}.
Based on symmetry arguments, such a rotor-embedded fluid layer can be viewed as a 2D chiral active fluid.
These molecular spinners can break time-reversal and parity symmetries within the fluid membrane~\cite{banerjee2017}.

The ES framework can also be generalized to fluids 
whose constitutive relations are not those of ordinary passive Newtonian fluid membranes.
When a fluid layer possesses chirality around an axis normal to its surface, an additional viscosity coefficient 
can emerge, even under the assumption of in-plane isotropy~\cite{avron1998}.
This isotropy-compatible transport coefficient is called odd viscosity, and it directly reflects the parity-violating nature of the chiral active fluid.
This symmetry breaking introduces an additional odd-viscous stress, which gives rise to a transverse velocity response in the 
force-balance equation
\begin{align}
	\eta_{\rm m} \nabla^2\mathbf{v} + 
		\eta_{\rm o} \bm\epsilon\cdot \nabla^2\mathbf{v}
	- \nabla p - \frac{\eta}{h} \mathbf{v} = \mathbf{0},
	\label{eq:ES_stokesodd}
\end{align}
where $\eta_{\rm o}$ is the odd viscosity coefficient and $\bm\epsilon$ denotes the 2D 
Levi-Civita tensor ($\epsilon_{xx}=\epsilon_{yy}=0$ and $\epsilon_{xy}=-\epsilon_{yx}=1$).

Because the antisymmetric tensor $\bm\epsilon$ represents an in-plane rotation of $\pi/2$, 
the odd viscosity contributes to an interfacial stress that is perpendicular to the one generated by the conventional shear 
viscosity $\eta_{\rm m}$.
We note that $\eta_{\rm o}$ is a dissipationless coefficient and is not constrained by the positivity of the energy dissipation.
Hence, $\eta_{\rm o}$ can take either positive or negative values depending on the fluid chirality.
This property is in contrast to shear viscosity, which must remain strictly positive, because it is associated with energy dissipation in the fluid.

Despite its potential importance as a new measure to estimate chirality in fluid membranes, odd viscosity is hard to measure in 2D~\cite{ganeshan2017}.
Its elusive nature stems from the fact that the odd term in the Stokes equation can be absorbed into the pressure, rendering 
the velocity field independent of $\eta_{\rm o}$ under conventional boundary conditions.
However, this argument is invalid when the odd viscosity is spatially nonuniform, as we discuss next.

\subsection{Inhomogeneous odd viscosity}
\label{sec:oddsub1}

To characterize 2D transport induced by odd viscosity,
Hosaka \textit{et al.}\ studied the hydrodynamic flow field and the force acting on a 2D liquid domain
in a supported membrane [see Fig.~\ref{fig:system}(c)]~\cite{hosaka2021hydrodynamic}.
In addition to the conventional membrane shear viscosity $\eta_{\rm m}$,
they specifically considered odd viscosities outside and inside the domain, denoted by $\eta_{\rm o}$ and $\eta_{\rm o}^\prime$, respectively. 
For simplicity, we hereafter set the inside and outside shear viscosities equal: $E=\eta_{\rm m}/\eta_{\rm m}^\prime=1$.

They first investigated the fluid flow induced by the lateral translational motion of a liquid domain in the presence of odd viscosities.
In Fig.~\ref{fig:3}, the velocity field in the comoving frame,  $\mathbf{v}-\mathbf{V}$, 
is plotted for (a) $\eta_{\rm o}=\eta_{\rm o}^\prime=\eta_{\rm m}$ (uniform odd viscosity),
(b) $\eta_{\rm o}=\eta_{\rm m}$ and $\eta_{\rm o}^\prime=0$ (vanishing odd viscosity inside the domain), 
(c) $\eta_{\rm o}=0$ and $\eta_{\rm o}^\prime=\eta_{\rm m}$ (vanishing odd viscosity outside the domain), 
and (d) $\eta_{\rm o}=-\eta_{\rm o}^\prime=\eta_{\rm m}$ (positive odd viscosity outside the domain and negative odd viscosity inside the domain).
In these plots, the domain size is fixed at $\kappa a = 0.1$.

When the odd viscosity is spatially uniform $(\eta_{\rm o}=\eta_{\rm o}^\prime)$, as in Fig.~\ref{fig:3}(a),  
the streamlines induced by the domain motion are symmetric with respect to the $x$-axis, 
which is the direction of motion of the liquid domain.
Such a symmetric profile is also observed in the passive case, where odd viscosity does not exist~\cite{ramachandran2010drag}.
When $\eta_{\rm o}\neq\eta_{\rm o}^\prime$, as in Figs.~\ref{fig:3}(b) and (c), 
the flow inside the domain rotates with respect to the $x$-axis and the symmetry is broken.
When $\eta_{\rm o}/\eta_{\rm o}^\prime<0$, as in Fig.~\ref{fig:3}(d), the flow inside the domain rotates more strongly,
as compared to Figs.~\ref{fig:3}(b) and (c).
Figure~\ref{fig:3}(c) is relevant to a lipid domain enriched with active rotor proteins, while Fig.~\ref{fig:3}(d) 
represents active proteins rotating oppositely inside and outside the domain.

\subsection{Hydrodynamic lift of a liquid domain}
\label{sec:oddsub2}

By inserting the velocity fields into the stress tensor and integrating it over the perimeter of the liquid domain, 
one can calculate the net force acting on it.
The rescaled drag coefficient along the $x$-axis is obtained as~\cite{hosaka2021hydrodynamic}
\begin{align}
\frac{\zeta_{xx}}{4\pi\eta_{\rm m}}&=
\frac{(\kappa a)^2}{4}
\nonumber \\
& +\frac{\kappa a K_1}{K_0}
\left[1- \frac{(\kappa a)^2(K_0I_1+K_1I_2)K_1I_2}
{[\kappa a(K_0I_1+K_1I_2)]^2+4(\delta K_0I_2)^2}
\right],
\label{eq:drag}
\end{align}
where $\delta=(\eta_{\rm o}-\eta_{\rm o}^\prime)/\eta_{\rm m}$ denotes the dimensionless difference in the odd viscosities.
In contrast to the passive case without odd viscosity, a force component appears along the $y$-direction.
The corresponding lift coefficient is given by~\cite{hosaka2021hydrodynamic}
\begin{align}
\frac{\zeta_{yx}}{4\pi\eta_{\rm m}}&=
\frac{2\delta(\kappa a K_1I_2)^2}
{[\kappa a (K_0I_1+K_1I_2)]^2+4(\delta K_0I_2)^2}.
\label{eq:lift}
\end{align}

The moving domain experiences not only a conventional drag force opposing the motion ($\zeta_{xx}$),
but also a hydrodynamic lift force perpendicular to the motion direction ($\zeta_{yx}$).
This lift arises only if there is a chirality contrast between the domain and the surrounding environment, 
i.e., $\delta\neq0$.
When the odd viscosities are uniform in space and $\delta=0$, the lift coefficient $\zeta_{yx}$ 
vanishes and the drag coefficient $\zeta_{xx}$ reduces to that for a regular liquid domain, as given by Eq.~\eqref{frictionb} with $E=1$.
Since active rotating proteins can accumulate inside the domain, a difference in odd viscosity can arise between the inside 
and outside the liquid domain.
In such a case, a lateral lift force can arise in a direction perpendicular to the drag.

In vector notation, the force-velocity relation is expressed as $\mathbf{F}=-\bm\zeta\cdot\mathbf{V}$, 
where the $2\times2$ drag matrix is given by~\cite{hosaka2021hydrodynamic}
\begin{align}
	\bm\zeta = \zeta_{xx}\mathbf{I} - \zeta_{yx} \bm\epsilon.
\end{align}
Due to the antisymmetric contribution from the transverse drag, the standard reciprocity under transposition no longer holds, 
namely, $\zeta_{ij}\neq\zeta_{ji}$.
Since the resistance tensor for objects of arbitrary shape in conventional passive fluids is required to be 
symmetric~\cite{masoud2019}, the measurement of a finite lift coefficient, $\zeta_{yx}$, provides 
a novel experimental method for probing chirality in 2D fluidic environments.

It is worth mentioning that despite this asymmetric structure, the resistance coefficients satisfy 
a generalized symmetry under the sign reversal of the odd-viscosity parameter $\delta$, such that $\zeta_{ij}(\delta) = \zeta_{ji}(-\delta)$.
This reciprocal relation is a manifestation of the Onsager-Casimir reciprocity~\cite{fruchart2023odd}, which is a direct consequence of the generalized Lorentz reciprocal theorem under certain boundary conditions in fluids with odd viscosity~\cite{hosaka2023lorentz}.

\section{Summary and outlook}

The Evans-Sackmann model has provided a fundamental framework for inferring membrane properties and has 
motivated extensive theoretical and experimental work over the past four decades.
Its lasting impact derives in large part from the conceptual simplicity and broad applicability of the ES framework 
to support and confine membrane geometries.
Owing to this generality, the framework has been extended to a wide range of problems, 
including correlated diffusion in supported biomembranes, boundary-induced active transport, and the dynamics of viscoelastic membranes. 
Although the ES model also underlies experimental approaches for probing protein dynamics, 
the present review has focused primarily on theoretical developments concerning transport under various 
boundary conditions, phase-separation dynamics, active diffusion, and the hydrodynamics of chiral active fluids.

In this review, we have mainly focused on translational transport 
and related collective dynamics, although the original ES model also accounted for the rigid disk's rotational motion. 
It was shown that, for small inclusions, the rotational drag coefficient scales as $a^2$ with the disk size. In contrast, 
in the friction-dominated regime, it exhibits a stronger algebraic dependence, scaling as $a^{3}$~\cite{evans1988}.
This stronger size dependence reflects the increased importance of substrate-induced momentum decay.
Building on these resistance laws, Suja \textit{et al.}~\cite{chandran2025hydrodynamics} 
have recently developed a predictive theory that 
allows the interleaflet friction to be estimated from the intrinsic membrane viscosity.

While we have mainly discussed isotropic inclusions, several studies have shown that anisotropic extended objects, 
such as rods, needles, and filaments, exhibit qualitatively different hydrodynamic 
resistance~\cite{Levine2004,Fischer2004,Manikantan_2024, shi2024drag}. 
For rod-like inclusions in membranes, the drag becomes anisotropic, with distinct longitudinal, transverse, and rotational components. 
In particular, for sufficiently long objects, the longitudinal resistance grows linearly with length up to a weak logarithmic correction, 
whereas the transverse and rotational resistances can show stronger algebraic growth~\cite{Levine2004}. 
Similar behavior has also been reported for needles in Langmuir monolayers and for filaments in supported bilayers~\cite{Fischer2004}. 
These results suggest that, for anisotropic inclusions, ES-type momentum screening does not lead to a single scalar drag 
coefficient but to direction-dependent transport laws that depend sensitively on geometry.

Looking ahead, several open problems remain for supported membrane hydrodynamics.
On the theoretical side, it will be important to extend the ES framework to account for nonlinear effects, viscoelasticity, 
and strongly heterogeneous environments.
On the experimental side, advances in high-resolution imaging and microfabrication are expected to provide more quantitative 
tests of hydrodynamic screening and collective dynamics in complex membrane systems.
In particular, the interplay between activity, chirality, and confinement may reveal new mechanisms of transport and self-organization 
that go beyond the current linear-response description~\cite{hosaka2022nonequilibrium}.

\section*{Author contributions}
Conceptualization: all authors; investigation: all authors; writing -- original draft: all authors; writing -- review \& editing: all authors.

\section*{Conflicts of interest}
There are no conflicts to declare.

\section*{Data availability}
No new data were created or analyzed in this study. Data sharing does not apply to this article, as
this manuscript is a Review Article based exclusively on previously published literature.

\begin{acknowledgments}

This article is dedicated to the memory of Erich Sackmann, whose pioneering contributions to molecular, cellular, and membrane biophysics shaped 
and inspired generations of theoreticians and experimentalists. At a time when the interface between physics and biology was still viewed with skepticism 
by many in both disciplines, Erich had the vision and courage to pursue fundamentally interdisciplinary questions. Working in those early days often without the 
full support or recognition of either community, he helped establish the physical foundations of modern biophysics of the cell, and demonstrated 
the profound power of bringing physical principles to living systems.

Beyond his extraordinary scientific achievements, Erich embodied the very best qualities of scholarship: intellectual depth, generosity, 
curiosity, and humanity. He was a true gentleman of science and a lasting source of inspiration for our studies of biological and model membranes.

One of us (D.A.) had the privilege of visiting Erich's group in Munich over nearly four decades, participating in many of the {\it Winterschulen}, 
and benefiting immensely from Erich's insight, encouragement, and friendship. His scientific vision, personal warmth, and unwavering commitment 
to interdisciplinary science left an enduring mark on all who had the fortune to know and work with him.

We thank H.\ Diamant, M.\ Imai, S.\ Ramachandran, and K.\ Seki for useful discussions.
We also thank Z.\ Xiong for carefully reading the manuscript.
Y.H.\ acknowledges support from the Japan Science and Technology Agency (JST) CREST (Grant No.\ JPMJCR25Q1).
D.A.\  acknowledges partial support from the Israel Science Foundation (ISF) under grant No. 226/24.
S.K.\ acknowledges support from the National Natural Science Foundation of China (Grant No.\ 12274098) 
and from the Zhejiang Key Laboratory of Soft Matter Biomedical Materials (2025ZY01036 and 2025E10072).
This work was supported by the JSPS Core-to-Core Program ``Advanced core-to-core network 
for the physics of self-organizing active matter" (JPJSCCA20230002).

\end{acknowledgments}

\appendix
\section{\\Derivation of the translational drag coefficient, Eq.~\eqref{eq:ES}}
\label{app}

We present the derivation of the translational drag coefficient for a disk moving 
in a 2D fluid membrane supported by a rigid substrate~\cite{evans1988}.
According to the Helmholtz decomposition, any 2D fluid velocity can be expressed as a gradient of a  
scalar potential $\varphi$ and the curl of a vector $\mathbf{A}= (0,0,A)$ that has only a $z$-component.
Then, the velocity is expressed as~\cite{ramachandran2010drag}
\begin{equation}
\mathbf{v}= -\nabla \varphi + \nabla \times \mathbf{A}.
\label{hmeqn:10}
\end{equation}
Taking the 2D curl of Eq.~\eqref{eq:ES_stokes} gives the Helmholtz equation for $A$
\begin{align}
(\nabla^2-\kappa^2)A=0,
\label{eq:helmholtz}
\end{align}
while $\varphi$ satisfies the Laplace equation
\begin{align}
\nabla^2\varphi=0.
\label{eq:laplace}
\end{align}
The hydrodynamic problem reduces to solving Eqs.~\eqref{eq:helmholtz} and \eqref{eq:laplace} 
under appropriate boundary conditions on the moving object.

Without loss of generality, we consider a 2D object with lateral translational velocity along the $x$-direction.
In the laboratory frame, the boundary condition for this motion is $\mathbf{v} = \mathbf{0}$ 
for $r\to\infty$. 
It is convenient to work in the 2D polar coordinates $(r,\theta)$ with the origin at the center of the circular disk.
For a disk of radius $a$ moving with $\mathbf{V}=(-V,0)$, the no-slip boundary condition at $r=a$ is
\begin{align}
v_r =- V \cos\theta, \quad
v_\theta =V\sin\theta.
\label{eq:bc2}
\end{align}
Under these boundary conditions, one can write the solutions to Eqs.~\eqref{eq:helmholtz} and \eqref{eq:laplace} as
\begin{align}
\varphi = \frac{C_1}{r}\cos\theta , \quad
A = C_2K_1(\kappa r)\sin\theta,
\label{solutionout}
\end{align}
where the coefficients are given by
\begin{align}
C_1 = -Va^2\left(1+\frac{2K_1(\kappa a)}{\kappa a K_0(\kappa a)}\right),\quad
C_2 = \frac{2V}{\kappa K_0(\kappa a)}.
\end{align}
Here, $K_n(z)$ are the modified Bessel functions of the second kind of order $n$.

By inserting the above solutions into Eq.~\eqref{hmeqn:10}, the radial and tangential 
components of the velocity are obtained as~\cite{ramachandran2010drag} 
\begin{align}
v_r&=\left[\frac{C_1}{r^2} + \frac{C_2}{r}K_1(\kappa r)
\right] \cos\theta, \\
v_\theta&=\left[
\frac{C_1}{r^2} + C_2 \kappa K_0(\kappa r) + 
\frac{C_2}{r}K_1(\kappa r)
\right] \sin\theta.
\end{align}
Then, the components of the stress tensor 
$\bm\sigma=-p\mathbf{I}+\eta_{\rm m}[\nabla\mathbf{v}+(\nabla\mathbf{v})^\top]$ 
($\top$ indicates the transpose) become
\begin{align}
\sigma_{rr} & =
- \eta_{\rm m}\left[
C_1 \left( 
\frac{\kappa^2}{r} 
+\frac{4}{r^3} \right) 
\right. 
\nonumber \\
& +  \left. 
C_2 \left( \frac{4}{r^2} K_1(\kappa r)
+ \frac{2\kappa}{r} K_0(\kappa r) \right) \right]
\cos\theta, 
\label{eq:srr}
\\
\sigma_{r\theta} & = 
 -\eta_{\rm m}\left[ 
\frac{4C_1}{r^3} 
\right. 
\nonumber \\ 
& + \left. 
C_2 \left( 
\frac{4}{r^2}K_1(\kappa r)  
+\kappa^2   K_1 (\kappa r)
+\frac{2 \kappa}{r} K_0(\kappa r)
\right)\right] \sin\theta.
\label{eq:srt}
\end{align}

To evaluate the net force exerted on the moving disk, we calculate the following integrals of the stress tensor over the 
disk perimeter:
\begin{align}
F_x&= a \int_0^{2\pi} {\rm d}\theta \, 
(\sigma_{rr} \cos\theta - 
\sigma_{r\theta}\sin \theta),\\
F_y&= a \int_0^{2\pi} {\rm d}\theta \, 
(\sigma_{rr} \sin\theta + 
\sigma_{r\theta}\cos \theta).
\end{align}
For a no-slip disk in a supported membrane, the perpendicular force vanishes, $F_y=0$.
From the linear relation, $\zeta=F_x/V$, the drag coefficient becomes 
\begin{equation}
\frac{\zeta}{4\pi\eta_{\rm m}}= 
\frac{(\kappa a)^2}{4} + 
\frac{\kappa a K_1(\kappa a)}{K_0(\kappa a)},
\end{equation}
which is Eq.~\eqref{eq:ES}.

\section{\\Derivation of the mobility tensor, Eq.~\eqref{eq:mobility_es}}
\label{appb}

To derive the mobility tensor of a supported membrane, we start with the ES hydrodynamic 
equations~\eqref{eq:ES_stokes} and \eqref{eq:incomp}, 
\begin{align}
	\eta_{\rm m} \nabla^2\mathbf{v} - \nabla p - \frac{\eta}{h} \mathbf{v} = - \mathbf{f},
	\quad 
	\nabla \cdot \mathbf{v}=0,
	\label{apphydroeq}
\end{align}
where we have introduced the 2D external force $\mathbf{f}(\mathbf{r})$.
We define the 2D Fourier transform as 
\begin{align}
{\mathbf v}({\mathbf r}) = 
\int \frac{{\rm d} \mathbf k}{(2\pi)^2}
{\mathbf v}_{\mathbf k}
\exp ( i {\mathbf k}\cdot {\mathbf r}),
\label{eqn:FT}
\end{align}
where $\mathbf k=(k_x, k_y)$ is the 2D wave vector.
Then, Eq.~\eqref{apphydroeq} can be rewritten as 
\begin{align}
	\eta_{\rm m} k^2 \mathbf{v}_{\mathbf k} + i \mathbf k p_{\mathbf k} + \frac{\eta}{h} \mathbf{v}_{\mathbf k} = \mathbf{f}_{\mathbf k},
	\quad 
	\mathbf k \cdot \mathbf{v}_{\mathbf k}=0,
\end{align}
where $k=\vert \mathbf k \vert$.
Hence, we obtain the velocity and pressure fields in Fourier space as
\begin{align}
	\mathbf{v}_{\mathbf k}&=\frac{1}{\eta_{\rm m}(k^2+\kappa^2)} \left( \mathbf{I} - \frac{\mathbf{k\,k}}{k^2} \right)
	\cdot \mathbf{f}_{\mathbf k},\\
	\label{fourierv}
	p_{\mathbf k} &=-i\frac{\mathbf{k}}{k^2}\cdot \mathbf{f}_{\mathbf k},
\end{align}
where we have used Eq.~\eqref{screeninglength} for $\kappa$. 
Then, going back to real space, we have  
\begin{align}
	\mathbf{v}(\mathbf r)=\int {\rm d} \mathbf{r}' \, \mathbf{G}(\mathbf{r}-\mathbf{r}') \cdot \mathbf{f}(\mathbf{r}'),\label{vr}
\end{align}
where 
\begin{align}
	\mathbf{G}(\mathbf{r}) = \int \frac{{\rm d}\mathbf k}{(2\pi)^2}
	\frac{1}{\eta_{\rm m}(k^2+\kappa^2)} \left( \mathbf{I} - \frac{\mathbf{k\,k}}{k^2} \right) 
	\exp ( i {\mathbf k}\cdot {\mathbf r}). 
	\label{Gr}
\end{align}
The pressure field is given by
\begin{align}
	p(\mathbf r)=
	\frac{1}{2\pi}
	\int {\rm d} \mathbf{r}' \, 
	\frac{\mathbf{r}-\mathbf{r}'}{\vert \mathbf{r}-\mathbf{r}' \vert^2}
	\cdot \mathbf{f}(\mathbf{r}').
\end{align}

Since the tensor $\mathbf{G}(\mathbf{r})$ depends only on the position vector $\mathbf{r}$, it can be written as 
\begin{equation}
\mathbf{G}({\mathbf{r}}) = 
c_1 \mathbf{I} + c_2 \frac{\mathbf{r\,r}}{r^2},
\label{eqn:realGES}
\end{equation}
with two coefficients $c_1$ and $c_2$.
By considering the trace and longitudinal component of Eq.~\eqref{eqn:realGES}, we obtain
\begin{align}
2c_1+c_2 &=
\frac{1}{2\pi\eta_{\rm m}} 
\int_0^\infty {\rm d}k \,
\frac{k J_0(kr)}{k^2 + \kappa^2}
\nonumber \\
&= \frac{1}{2\pi\eta_{\rm m}} K_0(\kappa r),
\label{eqn:es2AB}
\end{align}
and 
\begin{align}
c_1+c_2 
&= \frac{1}{2\pi\eta_{\rm m}}
\int_0^\infty {\rm d} k \,
\frac{J_1(kr)}{r(k^2 + \kappa^2)}
\nonumber \\
&= \frac{1}{2\pi\eta_{\rm m}} 
\left[ \frac{1}{(\kappa r)^2} - \frac{K_1(\kappa r)}{\kappa r} \right],
\label{eqn:esAB}
\end{align}
where $J_n(z)$ are Bessel functions of the first kind of order $n$.
Solving these equations, we obtain 
\begin{align}
c_1 & = 
\frac{1}{2\pi\eta_{\rm m}} \left[ K_0(\kappa r) 
+ \frac{K_1(\kappa r)}{\kappa r} 
- \frac{1}{(\kappa r)^2} \right],\\
c_2 &= \frac{1}{2\pi\eta_{\rm m}} \left[ -K_0(\kappa r) 
- \frac{2K_1(\kappa r)}{\kappa r} 
+ \frac{2}{(\kappa r)^2} \right].
\end{align}
Hence, we obtain Eq.~\eqref{eq:mobility_es}.

%


\begin{thebibliography}{88}%
\makeatletter
\providecommand \@ifxundefined [1]{%
 \@ifx{#1\undefined}
}%
\providecommand \@ifnum [1]{%
 \ifnum #1\expandafter \@firstoftwo
 \else \expandafter \@secondoftwo
 \fi
}%
\providecommand \@ifx [1]{%
 \ifx #1\expandafter \@firstoftwo
 \else \expandafter \@secondoftwo
 \fi
}%
\providecommand \natexlab [1]{#1}%
\providecommand \enquote  [1]{``#1''}%
\providecommand \bibnamefont  [1]{#1}%
\providecommand \bibfnamefont [1]{#1}%
\providecommand \citenamefont [1]{#1}%
\providecommand \href@noop [0]{\@secondoftwo}%
\providecommand \href [0]{\begingroup \@sanitize@url \@href}%
\providecommand \@href[1]{\@@startlink{#1}\@@href}%
\providecommand \@@href[1]{\endgroup#1\@@endlink}%
\providecommand \@sanitize@url [0]{\catcode `\\12\catcode `\$12\catcode
  `\&12\catcode `\#12\catcode `\^12\catcode `\_12\catcode `\%12\relax}%
\providecommand \@@startlink[1]{}%
\providecommand \@@endlink[0]{}%
\providecommand \url  [0]{\begingroup\@sanitize@url \@url }%
\providecommand \@url [1]{\endgroup\@href {#1}{\urlprefix }}%
\providecommand \urlprefix  [0]{URL }%
\providecommand \Eprint [0]{\href }%
\providecommand \doibase [0]{https://doi.org/}%
\providecommand \selectlanguage [0]{\@gobble}%
\providecommand \bibinfo  [0]{\@secondoftwo}%
\providecommand \bibfield  [0]{\@secondoftwo}%
\providecommand \translation [1]{[#1]}%
\providecommand \BibitemOpen [0]{}%
\providecommand \bibitemStop [0]{}%
\providecommand \bibitemNoStop [0]{.\EOS\space}%
\providecommand \EOS [0]{\spacefactor3000\relax}%
\providecommand \BibitemShut  [1]{\csname bibitem#1\endcsname}%
\let\auto@bib@innerbib\@empty
\bibitem [{\citenamefont {Singer}\ and\ \citenamefont
  {Nicolson}(1972)}]{singer1972fluid}%
  \BibitemOpen
  \bibfield  {author} {\bibinfo {author} {\bibfnamefont {S.~J.}\ \bibnamefont
  {Singer}}\ and\ \bibinfo {author} {\bibfnamefont {G.~L.}\ \bibnamefont
  {Nicolson}},\ }\bibfield  {title} {\bibinfo {title} {The fluid mosaic model
  of the structure of cell membranes: Cell membranes are viewed as
  two-dimensional solutions of oriented globular proteins and lipids},\ }\href
  {https://doi.org/10.1126/science.175.4023.720} {\bibfield  {journal}
  {\bibinfo  {journal} {Science}\ }\textbf {\bibinfo {volume} {175}},\ \bibinfo
  {pages} {720} (\bibinfo {year} {1972})}\BibitemShut {NoStop}%
\bibitem [{\citenamefont {Alberts}\ \emph {et~al.}(2008)\citenamefont
  {Alberts}, \citenamefont {Johnson}, \citenamefont {Lewis}, \citenamefont
  {Raff}, \citenamefont {Roberts},\ and\ \citenamefont {Walter}}]{albertsbook}%
  \BibitemOpen
  \bibfield  {author} {\bibinfo {author} {\bibfnamefont {B.}~\bibnamefont
  {Alberts}}, \bibinfo {author} {\bibfnamefont {A.}~\bibnamefont {Johnson}},
  \bibinfo {author} {\bibfnamefont {J.}~\bibnamefont {Lewis}}, \bibinfo
  {author} {\bibfnamefont {M.}~\bibnamefont {Raff}}, \bibinfo {author}
  {\bibfnamefont {K.}~\bibnamefont {Roberts}},\ and\ \bibinfo {author}
  {\bibfnamefont {P.}~\bibnamefont {Walter}},\ }\href@noop {} {\emph {\bibinfo
  {title} {Mol. Biol. Cell}}}\ (\bibinfo  {publisher} {Garland Science},\
  \bibinfo {address} {New York},\ \bibinfo {year} {2008})\BibitemShut {NoStop}%
\bibitem [{\citenamefont {Edidin}\ \emph {et~al.}(1976)\citenamefont {Edidin},
  \citenamefont {Zagyansky},\ and\ \citenamefont
  {Lardner}}]{edidin1976measurement}%
  \BibitemOpen
  \bibfield  {author} {\bibinfo {author} {\bibfnamefont {M.}~\bibnamefont
  {Edidin}}, \bibinfo {author} {\bibfnamefont {Y.}~\bibnamefont {Zagyansky}},\
  and\ \bibinfo {author} {\bibfnamefont {T.}~\bibnamefont {Lardner}},\
  }\bibfield  {title} {\bibinfo {title} {Measurement of membrane protein
  lateral diffusion in single cells},\ }\href@noop {} {\bibfield  {journal}
  {\bibinfo  {journal} {Science}\ }\textbf {\bibinfo {volume} {191}},\ \bibinfo
  {pages} {466} (\bibinfo {year} {1976})}\BibitemShut {NoStop}%
\bibitem [{\citenamefont {Simons}\ and\ \citenamefont
  {Ikonen}(1997)}]{simons1997functional}%
  \BibitemOpen
  \bibfield  {author} {\bibinfo {author} {\bibfnamefont {K.}~\bibnamefont
  {Simons}}\ and\ \bibinfo {author} {\bibfnamefont {E.}~\bibnamefont
  {Ikonen}},\ }\bibfield  {title} {\bibinfo {title} {Functional rafts in cell
  membranes},\ }\href {https://doi.org/https://doi.org/10.1038/42408}
  {\bibfield  {journal} {\bibinfo  {journal} {Nature}\ }\textbf {\bibinfo
  {volume} {387}},\ \bibinfo {pages} {569} (\bibinfo {year}
  {1997})}\BibitemShut {NoStop}%
\bibitem [{\citenamefont {Kusumi}\ \emph {et~al.}(2012)\citenamefont {Kusumi},
  \citenamefont {Fujiwara}, \citenamefont {Chadda}, \citenamefont {Xie},
  \citenamefont {Tsunoyama}, \citenamefont {Kalay}, \citenamefont {Kasai},\
  and\ \citenamefont {Suzuki}}]{kusumi2012dynamic}%
  \BibitemOpen
  \bibfield  {author} {\bibinfo {author} {\bibfnamefont {A.}~\bibnamefont
  {Kusumi}}, \bibinfo {author} {\bibfnamefont {T.~K.}\ \bibnamefont
  {Fujiwara}}, \bibinfo {author} {\bibfnamefont {R.}~\bibnamefont {Chadda}},
  \bibinfo {author} {\bibfnamefont {M.}~\bibnamefont {Xie}}, \bibinfo {author}
  {\bibfnamefont {T.~A.}\ \bibnamefont {Tsunoyama}}, \bibinfo {author}
  {\bibfnamefont {Z.}~\bibnamefont {Kalay}}, \bibinfo {author} {\bibfnamefont
  {R.~S.}\ \bibnamefont {Kasai}},\ and\ \bibinfo {author} {\bibfnamefont
  {K.~G.}\ \bibnamefont {Suzuki}},\ }\bibfield  {title} {\bibinfo {title}
  {{Dynamic organizing principles of the plasma membrane that regulate signal
  transduction: Commemorating the fortieth anniversary of Singer and Nicolson's
  fluid-mosaic model}},\ }\href
  {https://doi.org/10.1146/annurev-cellbio-100809-151736} {\bibfield  {journal}
  {\bibinfo  {journal} {Annu. Rev. Cell Dev. Biol.}\ }\textbf {\bibinfo
  {volume} {28}},\ \bibinfo {pages} {215} (\bibinfo {year} {2012})}\BibitemShut
  {NoStop}%
\bibitem [{\citenamefont {Landau}\ and\ \citenamefont
  {Lifshitz}(1987)}]{Landau1987}%
  \BibitemOpen
  \bibfield  {author} {\bibinfo {author} {\bibfnamefont {L.~D.}\ \bibnamefont
  {Landau}}\ and\ \bibinfo {author} {\bibfnamefont {E.~M.}\ \bibnamefont
  {Lifshitz}},\ }\href@noop {} {\emph {\bibinfo {title} {Fluid Mechanics}}},\
  Vol.~\bibinfo {volume} {6}\ (\bibinfo  {publisher} {Pergamon Press},\
  \bibinfo {address} {Oxford},\ \bibinfo {year} {1987})\BibitemShut {NoStop}%
\bibitem [{\citenamefont {Happel}\ and\ \citenamefont
  {Brenner}(1983)}]{happel2012low}%
  \BibitemOpen
  \bibfield  {author} {\bibinfo {author} {\bibfnamefont {J.}~\bibnamefont
  {Happel}}\ and\ \bibinfo {author} {\bibfnamefont {H.}~\bibnamefont
  {Brenner}},\ }\href@noop {} {\emph {\bibinfo {title} {{Low Reynolds Number
  Hydrodynamics}}}}\ (\bibinfo  {publisher} {Springer},\ \bibinfo {address}
  {Netherlands},\ \bibinfo {year} {1983})\BibitemShut {NoStop}%
\bibitem [{\citenamefont {Lipowsky}\ and\ \citenamefont
  {Sackmann}(1995)}]{lipowsky1995structure}%
  \BibitemOpen
  \bibfield  {author} {\bibinfo {author} {\bibfnamefont {R.}~\bibnamefont
  {Lipowsky}}\ and\ \bibinfo {author} {\bibfnamefont {E.}~\bibnamefont
  {Sackmann}},\ }\href@noop {} {\emph {\bibinfo {title} {Structure and dynamics
  of membranes: I. from cells to vesicles/II. generic and specific
  interactions}}}\ (\bibinfo  {publisher} {Elsevier},\ \bibinfo {year}
  {1995})\BibitemShut {NoStop}%
\bibitem [{\citenamefont {Saffman}\ and\ \citenamefont
  {Delbr{\"u}ck}(1975)}]{saffman1975}%
  \BibitemOpen
  \bibfield  {author} {\bibinfo {author} {\bibfnamefont {P.~G.}\ \bibnamefont
  {Saffman}}\ and\ \bibinfo {author} {\bibfnamefont {M.}~\bibnamefont
  {Delbr{\"u}ck}},\ }\bibfield  {title} {\bibinfo {title} {Brownian motion in
  biological membranes},\ }\href
  {https://doi.org/https://doi.org/10.1073/pnas.72.8.3111} {\bibfield
  {journal} {\bibinfo  {journal} {Proc. Natl. Acad. Sci. U.S.A.}\ }\textbf
  {\bibinfo {volume} {72}},\ \bibinfo {pages} {3111} (\bibinfo {year}
  {1975})}\BibitemShut {NoStop}%
\bibitem [{\citenamefont {Saffman}(1976)}]{saffman1976}%
  \BibitemOpen
  \bibfield  {author} {\bibinfo {author} {\bibfnamefont {P.~G.}\ \bibnamefont
  {Saffman}},\ }\bibfield  {title} {\bibinfo {title} {Brownian motion in thin
  sheets of viscous fluid},\ }\href
  {https://doi.org/https://doi.org/10.1017/S0022112076001511} {\bibfield
  {journal} {\bibinfo  {journal} {J. Fluid Mech.}\ }\textbf {\bibinfo {volume}
  {73}},\ \bibinfo {pages} {593} (\bibinfo {year} {1976})}\BibitemShut
  {NoStop}%
\bibitem [{\citenamefont {Hughes}\ \emph {et~al.}(1981)\citenamefont {Hughes},
  \citenamefont {Pailthorpe},\ and\ \citenamefont
  {White}}]{hughes1981translational}%
  \BibitemOpen
  \bibfield  {author} {\bibinfo {author} {\bibfnamefont {B.}~\bibnamefont
  {Hughes}}, \bibinfo {author} {\bibfnamefont {B.}~\bibnamefont {Pailthorpe}},\
  and\ \bibinfo {author} {\bibfnamefont {L.}~\bibnamefont {White}},\ }\bibfield
   {title} {\bibinfo {title} {The translational and rotational drag on a
  cylinder moving in a membrane},\ }\href@noop {} {\bibfield  {journal}
  {\bibinfo  {journal} {J. Fluid Mech.}\ }\textbf {\bibinfo {volume} {110}},\
  \bibinfo {pages} {349} (\bibinfo {year} {1981})}\BibitemShut {NoStop}%
\bibitem [{\citenamefont {Henle}\ and\ \citenamefont
  {Levine}(2010)}]{henle2010}%
  \BibitemOpen
  \bibfield  {author} {\bibinfo {author} {\bibfnamefont {M.~L.}\ \bibnamefont
  {Henle}}\ and\ \bibinfo {author} {\bibfnamefont {A.~J.}\ \bibnamefont
  {Levine}},\ }\bibfield  {title} {\bibinfo {title} {Hydrodynamics in curved
  membranes: The effect of geometry on particulate mobility},\ }\href@noop {}
  {\bibfield  {journal} {\bibinfo  {journal} {Phys. Rev. E}\ }\textbf {\bibinfo
  {volume} {81}},\ \bibinfo {pages} {011905} (\bibinfo {year}
  {2010})}\BibitemShut {NoStop}%
\bibitem [{\citenamefont {Shi}\ \emph {et~al.}(2024)\citenamefont {Shi},
  \citenamefont {Moradi},\ and\ \citenamefont {Nazockdast}}]{shi2024drag}%
  \BibitemOpen
  \bibfield  {author} {\bibinfo {author} {\bibfnamefont {W.}~\bibnamefont
  {Shi}}, \bibinfo {author} {\bibfnamefont {M.}~\bibnamefont {Moradi}},\ and\
  \bibinfo {author} {\bibfnamefont {E.}~\bibnamefont {Nazockdast}},\ }\bibfield
   {title} {\bibinfo {title} {The drag of a filament moving in a supported
  spherical bilayer},\ }\href@noop {} {\bibfield  {journal} {\bibinfo
  {journal} {J. Fluid Mech.}\ }\textbf {\bibinfo {volume} {979}},\ \bibinfo
  {pages} {A6} (\bibinfo {year} {2024})}\BibitemShut {NoStop}%
\bibitem [{\citenamefont {Oppenheimer}\ and\ \citenamefont
  {Diamant}(2009)}]{oppenheimer2009}%
  \BibitemOpen
  \bibfield  {author} {\bibinfo {author} {\bibfnamefont {N.}~\bibnamefont
  {Oppenheimer}}\ and\ \bibinfo {author} {\bibfnamefont {H.}~\bibnamefont
  {Diamant}},\ }\bibfield  {title} {\bibinfo {title} {Correlated diffusion of
  membrane proteins and their effect on membrane viscosity},\ }\href
  {https://doi.org/https://doi.org/10.1016/j.bpj.2009.01.020} {\bibfield
  {journal} {\bibinfo  {journal} {Biophys. J.}\ }\textbf {\bibinfo {volume}
  {96}},\ \bibinfo {pages} {3041} (\bibinfo {year} {2009})}\BibitemShut
  {NoStop}%
\bibitem [{\citenamefont {Sackmann}(1996)}]{sackmann1996supported}%
  \BibitemOpen
  \bibfield  {author} {\bibinfo {author} {\bibfnamefont {E.}~\bibnamefont
  {Sackmann}},\ }\bibfield  {title} {\bibinfo {title} {Supported membranes:
  Scientific and practical applications},\ }\href
  {https://doi.org/10.1126/science.271.5245.43} {\bibfield  {journal} {\bibinfo
   {journal} {Science}\ }\textbf {\bibinfo {volume} {271}},\ \bibinfo {pages}
  {43} (\bibinfo {year} {1996})}\BibitemShut {NoStop}%
\bibitem [{\citenamefont {Tanaka}\ and\ \citenamefont
  {Sackmann}(2005)}]{tanaka2005polymer}%
  \BibitemOpen
  \bibfield  {author} {\bibinfo {author} {\bibfnamefont {M.}~\bibnamefont
  {Tanaka}}\ and\ \bibinfo {author} {\bibfnamefont {E.}~\bibnamefont
  {Sackmann}},\ }\bibfield  {title} {\bibinfo {title} {Polymer-supported
  membranes as models of the cell surface},\ }\href@noop {} {\bibfield
  {journal} {\bibinfo  {journal} {Nature}\ }\textbf {\bibinfo {volume} {437}},\
  \bibinfo {pages} {656} (\bibinfo {year} {2005})}\BibitemShut {NoStop}%
\bibitem [{\citenamefont {Castellana}\ and\ \citenamefont
  {Cremer}(2006)}]{Castellana2006}%
  \BibitemOpen
  \bibfield  {author} {\bibinfo {author} {\bibfnamefont {E.~T.}\ \bibnamefont
  {Castellana}}\ and\ \bibinfo {author} {\bibfnamefont {P.~S.}\ \bibnamefont
  {Cremer}},\ }\bibfield  {title} {\bibinfo {title} {Solid supported lipid
  bilayers: From biophysical studies to sensor design},\ }\href@noop {}
  {\bibfield  {journal} {\bibinfo  {journal} {Surf. Sci. Rep.}\ }\textbf
  {\bibinfo {volume} {61}},\ \bibinfo {pages} {429} (\bibinfo {year}
  {2006})}\BibitemShut {NoStop}%
\bibitem [{\citenamefont {Ainla}\ \emph {et~al.}(2013)\citenamefont {Ainla},
  \citenamefont {Jansson}, \citenamefont {Stepanyants}, \citenamefont {Orwar},\
  and\ \citenamefont {Jesorka}}]{Ainla2013}%
  \BibitemOpen
  \bibfield  {author} {\bibinfo {author} {\bibfnamefont {A.}~\bibnamefont
  {Ainla}}, \bibinfo {author} {\bibfnamefont {E.~T.}\ \bibnamefont {Jansson}},
  \bibinfo {author} {\bibfnamefont {N.}~\bibnamefont {Stepanyants}}, \bibinfo
  {author} {\bibfnamefont {O.}~\bibnamefont {Orwar}},\ and\ \bibinfo {author}
  {\bibfnamefont {A.}~\bibnamefont {Jesorka}},\ }\bibfield  {title} {\bibinfo
  {title} {Lab on a biomembrane: {Rapid} prototyping and manipulation of {2D}
  fluidic lipid bilayer circuits},\ }\href@noop {} {\bibfield  {journal}
  {\bibinfo  {journal} {Sci. Rep.}\ }\textbf {\bibinfo {volume} {3}},\ \bibinfo
  {pages} {2743} (\bibinfo {year} {2013})}\BibitemShut {NoStop}%
\bibitem [{\citenamefont {Evans}\ and\ \citenamefont
  {Sackmann}(1988)}]{evans1988}%
  \BibitemOpen
  \bibfield  {author} {\bibinfo {author} {\bibfnamefont {E.}~\bibnamefont
  {Evans}}\ and\ \bibinfo {author} {\bibfnamefont {E.}~\bibnamefont
  {Sackmann}},\ }\bibfield  {title} {\bibinfo {title} {Translational and
  rotational drag coefficients for a disk moving in a liquid membrane
  associated with a rigid substrate},\ }\href
  {https://doi.org/10.1017/S0022112088003106} {\bibfield  {journal} {\bibinfo
  {journal} {J. Fluid Mech.}\ }\textbf {\bibinfo {volume} {194}},\ \bibinfo
  {pages} {553} (\bibinfo {year} {1988})}\BibitemShut {NoStop}%
\bibitem [{\citenamefont {Merkel}\ \emph {et~al.}(1989)\citenamefont {Merkel},
  \citenamefont {Sackmann},\ and\ \citenamefont {Evans}}]{merkel1989molecular}%
  \BibitemOpen
  \bibfield  {author} {\bibinfo {author} {\bibfnamefont {R.}~\bibnamefont
  {Merkel}}, \bibinfo {author} {\bibfnamefont {E.}~\bibnamefont {Sackmann}},\
  and\ \bibinfo {author} {\bibfnamefont {E.}~\bibnamefont {Evans}},\ }\bibfield
   {title} {\bibinfo {title} {Molecular friction and epitactic coupling between
  monolayers in supported bilayers},\ }\href@noop {} {\bibfield  {journal}
  {\bibinfo  {journal} {J. Phys. (Paris)}\ }\textbf {\bibinfo {volume} {50}},\
  \bibinfo {pages} {1535} (\bibinfo {year} {1989})}\BibitemShut {NoStop}%
\bibitem [{\citenamefont {Brinkman}(1949)}]{Brinkman1949}%
  \BibitemOpen
  \bibfield  {author} {\bibinfo {author} {\bibfnamefont {H.~C.}\ \bibnamefont
  {Brinkman}},\ }\bibfield  {title} {\bibinfo {title} {A calculation of the
  viscous force exerted by a flowing fluid on a dense swarm of particles},\
  }\href {https://doi.org/10.1007/BF02120313} {\bibfield  {journal} {\bibinfo
  {journal} {Appl. Sci. Res. A}\ }\textbf {\bibinfo {volume} {1}},\ \bibinfo
  {pages} {27} (\bibinfo {year} {1949})}\BibitemShut {NoStop}%
\bibitem [{\citenamefont {Sackmann}\ and\ \citenamefont
  {Tanaka}(2000)}]{sackmann2000supported}%
  \BibitemOpen
  \bibfield  {author} {\bibinfo {author} {\bibfnamefont {E.}~\bibnamefont
  {Sackmann}}\ and\ \bibinfo {author} {\bibfnamefont {M.}~\bibnamefont
  {Tanaka}},\ }\bibfield  {title} {\bibinfo {title} {Supported membranes on
  soft polymer cushions: fabrication, characterization and applications},\
  }\href@noop {} {\bibfield  {journal} {\bibinfo  {journal} {Trends
  Biotechnol.}\ }\textbf {\bibinfo {volume} {18}},\ \bibinfo {pages} {58}
  (\bibinfo {year} {2000})}\BibitemShut {NoStop}%
\bibitem [{\citenamefont {Stone}\ and\ \citenamefont
  {Ajdari}(1998)}]{stone1998hydrodynamics}%
  \BibitemOpen
  \bibfield  {author} {\bibinfo {author} {\bibfnamefont {H.~A.}\ \bibnamefont
  {Stone}}\ and\ \bibinfo {author} {\bibfnamefont {A.}~\bibnamefont {Ajdari}},\
  }\bibfield  {title} {\bibinfo {title} {Hydrodynamics of particles embedded in
  a flat surfactant layer overlying a subphase of finite depth},\ }\href@noop
  {} {\bibfield  {journal} {\bibinfo  {journal} {J. Fluid Mech.}\ }\textbf
  {\bibinfo {volume} {369}},\ \bibinfo {pages} {151} (\bibinfo {year}
  {1998})}\BibitemShut {NoStop}%
\bibitem [{\citenamefont {Camley}\ and\ \citenamefont
  {Brown}(2013)}]{camley2013diffusion}%
  \BibitemOpen
  \bibfield  {author} {\bibinfo {author} {\bibfnamefont {B.~A.}\ \bibnamefont
  {Camley}}\ and\ \bibinfo {author} {\bibfnamefont {F.~L.}\ \bibnamefont
  {Brown}},\ }\bibfield  {title} {\bibinfo {title} {Diffusion of complex
  objects embedded in free and supported lipid bilayer membranes: role of shape
  anisotropy and leaflet structure},\ }\href@noop {} {\bibfield  {journal}
  {\bibinfo  {journal} {Soft Matter}\ }\textbf {\bibinfo {volume} {9}},\
  \bibinfo {pages} {4767} (\bibinfo {year} {2013})}\BibitemShut {NoStop}%
\bibitem [{\citenamefont {Anthony}\ \emph {et~al.}(2022)\citenamefont
  {Anthony}, \citenamefont {Sahin}, \citenamefont {Yapici}, \citenamefont
  {Rogers},\ and\ \citenamefont {Honerkamp-Smith}}]{anthony2022systematic}%
  \BibitemOpen
  \bibfield  {author} {\bibinfo {author} {\bibfnamefont {A.~A.}\ \bibnamefont
  {Anthony}}, \bibinfo {author} {\bibfnamefont {O.}~\bibnamefont {Sahin}},
  \bibinfo {author} {\bibfnamefont {M.~K.}\ \bibnamefont {Yapici}}, \bibinfo
  {author} {\bibfnamefont {D.}~\bibnamefont {Rogers}},\ and\ \bibinfo {author}
  {\bibfnamefont {A.~R.}\ \bibnamefont {Honerkamp-Smith}},\ }\bibfield  {title}
  {\bibinfo {title} {Systematic measurements of interleaflet friction in
  supported bilayers},\ }\href {https://doi.org/10.1016/j.bpj.2022.06.023}
  {\bibfield  {journal} {\bibinfo  {journal} {Biophys. J.}\ }\textbf {\bibinfo
  {volume} {121}},\ \bibinfo {pages} {2981} (\bibinfo {year}
  {2022})}\BibitemShut {NoStop}%
\bibitem [{\citenamefont {Komura}\ \emph
  {et~al.}(2012{\natexlab{a}})\citenamefont {Komura}, \citenamefont
  {Ramachandran},\ and\ \citenamefont {Seki}}]{komura2012lateral}%
  \BibitemOpen
  \bibfield  {author} {\bibinfo {author} {\bibfnamefont {S.}~\bibnamefont
  {Komura}}, \bibinfo {author} {\bibfnamefont {S.}~\bibnamefont
  {Ramachandran}},\ and\ \bibinfo {author} {\bibfnamefont {K.}~\bibnamefont
  {Seki}},\ }\bibfield  {title} {\bibinfo {title} {Lateral dynamics in
  polymer-supported membranes},\ }\href
  {https://doi.org/https://doi.org/10.3390/ma5101923} {\bibfield  {journal}
  {\bibinfo  {journal} {Materials}\ }\textbf {\bibinfo {volume} {5}},\ \bibinfo
  {pages} {1923} (\bibinfo {year} {2012}{\natexlab{a}})}\BibitemShut {NoStop}%
\bibitem [{\citenamefont {Komura}\ \emph
  {et~al.}(2012{\natexlab{b}})\citenamefont {Komura}, \citenamefont
  {Ramachandran},\ and\ \citenamefont {Seki}}]{komura2012anomalous}%
  \BibitemOpen
  \bibfield  {author} {\bibinfo {author} {\bibfnamefont {S.}~\bibnamefont
  {Komura}}, \bibinfo {author} {\bibfnamefont {S.}~\bibnamefont
  {Ramachandran}},\ and\ \bibinfo {author} {\bibfnamefont {K.}~\bibnamefont
  {Seki}},\ }\bibfield  {title} {\bibinfo {title} {Anomalous lateral diffusion
  in a viscous membrane surrounded by viscoelastic media},\ }\href
  {https://doi.org/10.1209/0295-5075/97/68007} {\bibfield  {journal} {\bibinfo
  {journal} {EPL}\ }\textbf {\bibinfo {volume} {97}},\ \bibinfo {pages} {68007}
  (\bibinfo {year} {2012}{\natexlab{b}})}\BibitemShut {NoStop}%
\bibitem [{\citenamefont {Bar-Haim}\ and\ \citenamefont
  {Diamant}(2017)}]{bar2017correlations}%
  \BibitemOpen
  \bibfield  {author} {\bibinfo {author} {\bibfnamefont {C.}~\bibnamefont
  {Bar-Haim}}\ and\ \bibinfo {author} {\bibfnamefont {H.}~\bibnamefont
  {Diamant}},\ }\bibfield  {title} {\bibinfo {title} {Correlations in
  suspensions confined between viscoelastic surfaces: Noncontact
  microrheology},\ }\href
  {https://doi.org/https://doi.org/10.1103/PhysRevE.96.022607} {\bibfield
  {journal} {\bibinfo  {journal} {Phys. Rev. E}\ }\textbf {\bibinfo {volume}
  {96}},\ \bibinfo {pages} {022607} (\bibinfo {year} {2017})}\BibitemShut
  {NoStop}%
\bibitem [{\citenamefont {Abramowitz}\ and\ \citenamefont
  {Stegun}(2000)}]{abramowitz2000handbook}%
  \BibitemOpen
  \bibfield  {author} {\bibinfo {author} {\bibfnamefont {M.}~\bibnamefont
  {Abramowitz}}\ and\ \bibinfo {author} {\bibfnamefont {I.~A.}\ \bibnamefont
  {Stegun}},\ }\href@noop {} {\emph {\bibinfo {title} {{Handbook of
  Mathematical Functions with Formulas, Graphs, and Mathematical Tables}}}},\
  Vol.~\bibinfo {volume} {55}\ (\bibinfo  {publisher} {Dover Publications Inc,
  New York},\ \bibinfo {year} {2000})\BibitemShut {NoStop}%
\bibitem [{\citenamefont {Diamant}(2009)}]{diamant2009hydrodynamic}%
  \BibitemOpen
  \bibfield  {author} {\bibinfo {author} {\bibfnamefont {H.}~\bibnamefont
  {Diamant}},\ }\bibfield  {title} {\bibinfo {title} {Hydrodynamic interaction
  in confined geometries},\ }\href
  {https://doi.org/https://doi.org/10.1143/JPSJ.78.041002} {\bibfield
  {journal} {\bibinfo  {journal} {J. Phys. Soc. Jpn.}\ }\textbf {\bibinfo
  {volume} {78}},\ \bibinfo {pages} {041002} (\bibinfo {year}
  {2009})}\BibitemShut {NoStop}%
\bibitem [{\citenamefont {Doi}(2013)}]{doi2013soft}%
  \BibitemOpen
  \bibfield  {author} {\bibinfo {author} {\bibfnamefont {M.}~\bibnamefont
  {Doi}},\ }\href
  {https://doi.org/https://doi.org/10.1093/acprof:oso/9780199652952.001.0001}
  {\emph {\bibinfo {title} {Soft Matter Physics}}}\ (\bibinfo  {publisher}
  {Oxford University Press},\ \bibinfo {address} {New York},\ \bibinfo {year}
  {2013})\BibitemShut {NoStop}%
\bibitem [{\citenamefont {Seki}\ \emph {et~al.}(2014)\citenamefont {Seki},
  \citenamefont {Mogre},\ and\ \citenamefont {Komura}}]{seki2014diffusion}%
  \BibitemOpen
  \bibfield  {author} {\bibinfo {author} {\bibfnamefont {K.}~\bibnamefont
  {Seki}}, \bibinfo {author} {\bibfnamefont {S.}~\bibnamefont {Mogre}},\ and\
  \bibinfo {author} {\bibfnamefont {S.}~\bibnamefont {Komura}},\ }\bibfield
  {title} {\bibinfo {title} {Diffusion coefficients in leaflets of bilayer
  membranes},\ }\href
  {https://doi.org/https://doi.org/10.1103/PhysRevE.89.022713} {\bibfield
  {journal} {\bibinfo  {journal} {Phys. Rev. E}\ }\textbf {\bibinfo {volume}
  {89}},\ \bibinfo {pages} {022713} (\bibinfo {year} {2014})}\BibitemShut
  {NoStop}%
\bibitem [{\citenamefont {Hill}\ and\ \citenamefont
  {Wang}(2014)}]{hill2014diffusion}%
  \BibitemOpen
  \bibfield  {author} {\bibinfo {author} {\bibfnamefont {R.~J.}\ \bibnamefont
  {Hill}}\ and\ \bibinfo {author} {\bibfnamefont {C.-Y.}\ \bibnamefont
  {Wang}},\ }\bibfield  {title} {\bibinfo {title} {Diffusion in phospholipid
  bilayer membranes: Dual-leaflet dynamics and the roles of tracer-leaflet and
  inter-leaflet coupling},\ }\href {https://doi.org/10.1098/rspa.2013.0843}
  {\bibfield  {journal} {\bibinfo  {journal} {Proc. R. Soc. A}\ }\textbf
  {\bibinfo {volume} {470}},\ \bibinfo {pages} {20130843} (\bibinfo {year}
  {2014})}\BibitemShut {NoStop}%
\bibitem [{\citenamefont {Barentin}\ \emph {et~al.}(1999)\citenamefont
  {Barentin}, \citenamefont {Ybert}, \citenamefont {Di~Meglio},\ and\
  \citenamefont {Joanny}}]{barentin1999}%
  \BibitemOpen
  \bibfield  {author} {\bibinfo {author} {\bibfnamefont {C.}~\bibnamefont
  {Barentin}}, \bibinfo {author} {\bibfnamefont {C.}~\bibnamefont {Ybert}},
  \bibinfo {author} {\bibfnamefont {J.-M.}\ \bibnamefont {Di~Meglio}},\ and\
  \bibinfo {author} {\bibfnamefont {J.-F.}\ \bibnamefont {Joanny}},\ }\bibfield
   {title} {\bibinfo {title} {Surface shear viscosity of {G}ibbs and {L}angmuir
  monolayers},\ }\href {https://doi.org/10.1017/S0022112099006321} {\bibfield
  {journal} {\bibinfo  {journal} {J. Fluid Mech.}\ }\textbf {\bibinfo {volume}
  {397}},\ \bibinfo {pages} {331} (\bibinfo {year} {1999})}\BibitemShut
  {NoStop}%
\bibitem [{\citenamefont {Elfring}\ \emph {et~al.}(2016)\citenamefont
  {Elfring}, \citenamefont {Leal},\ and\ \citenamefont
  {Squires}}]{elfring2016surface}%
  \BibitemOpen
  \bibfield  {author} {\bibinfo {author} {\bibfnamefont {G.~J.}\ \bibnamefont
  {Elfring}}, \bibinfo {author} {\bibfnamefont {L.~G.}\ \bibnamefont {Leal}},\
  and\ \bibinfo {author} {\bibfnamefont {T.~M.}\ \bibnamefont {Squires}},\
  }\bibfield  {title} {\bibinfo {title} {Surface viscosity and {Marangoni}
  stresses at surfactant laden interfaces},\ }\href
  {https://doi.org/https://doi.org/10.1017/jfm.2016.96} {\bibfield  {journal}
  {\bibinfo  {journal} {J. Fluid Mech.}\ }\textbf {\bibinfo {volume} {792}},\
  \bibinfo {pages} {712} (\bibinfo {year} {2016})}\BibitemShut {NoStop}%
\bibitem [{\citenamefont {Manikantan}\ and\ \citenamefont
  {Squires}(2020)}]{manikantan2020surfactant}%
  \BibitemOpen
  \bibfield  {author} {\bibinfo {author} {\bibfnamefont {H.}~\bibnamefont
  {Manikantan}}\ and\ \bibinfo {author} {\bibfnamefont {T.~M.}\ \bibnamefont
  {Squires}},\ }\bibfield  {title} {\bibinfo {title} {Surfactant dynamics:
  hidden variables controlling fluid flows},\ }\href
  {https://doi.org/10.1017/jfm.2020.170} {\bibfield  {journal} {\bibinfo
  {journal} {J. Fluid Mech.}\ }\textbf {\bibinfo {volume} {892}},\ \bibinfo
  {pages} {P1} (\bibinfo {year} {2020})}\BibitemShut {NoStop}%
\bibitem [{\citenamefont {Hosaka}\ \emph
  {et~al.}(2021{\natexlab{a}})\citenamefont {Hosaka}, \citenamefont {Komura},\
  and\ \citenamefont {Andelman}}]{hosaka2021nonreciprocal}%
  \BibitemOpen
  \bibfield  {author} {\bibinfo {author} {\bibfnamefont {Y.}~\bibnamefont
  {Hosaka}}, \bibinfo {author} {\bibfnamefont {S.}~\bibnamefont {Komura}},\
  and\ \bibinfo {author} {\bibfnamefont {D.}~\bibnamefont {Andelman}},\
  }\bibfield  {title} {\bibinfo {title} {Nonreciprocal response of a
  two-dimensional fluid with odd viscosity},\ }\href
  {https://doi.org/10.1103/PhysRevE.103.042610} {\bibfield  {journal} {\bibinfo
   {journal} {Phys. Rev. E}\ }\textbf {\bibinfo {volume} {103}},\ \bibinfo
  {pages} {042610} (\bibinfo {year} {2021}{\natexlab{a}})}\BibitemShut
  {NoStop}%
\bibitem [{\citenamefont {Hosaka}\ \emph
  {et~al.}(2023{\natexlab{a}})\citenamefont {Hosaka}, \citenamefont
  {Andelman},\ and\ \citenamefont {Komura}}]{hosaka2023pair}%
  \BibitemOpen
  \bibfield  {author} {\bibinfo {author} {\bibfnamefont {Y.}~\bibnamefont
  {Hosaka}}, \bibinfo {author} {\bibfnamefont {D.}~\bibnamefont {Andelman}},\
  and\ \bibinfo {author} {\bibfnamefont {S.}~\bibnamefont {Komura}},\
  }\bibfield  {title} {\bibinfo {title} {Pair dynamics of active force dipoles
  in an odd-viscous fluid},\ }\href
  {https://doi.org/10.1140/epje/s10189-023-00265-y} {\bibfield  {journal}
  {\bibinfo  {journal} {Eur. Phys. J. E}\ }\textbf {\bibinfo {volume} {46}},\
  \bibinfo {pages} {18} (\bibinfo {year} {2023}{\natexlab{a}})}\BibitemShut
  {NoStop}%
\bibitem [{\citenamefont {Hosaka}\ \emph
  {et~al.}(2023{\natexlab{b}})\citenamefont {Hosaka}, \citenamefont
  {Golestanian},\ and\ \citenamefont
  {Daddi-Moussa-Ider}}]{hosaka2023hydrodynamics}%
  \BibitemOpen
  \bibfield  {author} {\bibinfo {author} {\bibfnamefont {Y.}~\bibnamefont
  {Hosaka}}, \bibinfo {author} {\bibfnamefont {R.}~\bibnamefont
  {Golestanian}},\ and\ \bibinfo {author} {\bibfnamefont {A.}~\bibnamefont
  {Daddi-Moussa-Ider}},\ }\bibfield  {title} {\bibinfo {title} {Hydrodynamics
  of an odd active surfer in a chiral fluid},\ }\href
  {https://doi.org/10.1088/1367-2630/aceea4} {\bibfield  {journal} {\bibinfo
  {journal} {New J. Phys.}\ }\textbf {\bibinfo {volume} {25}},\ \bibinfo
  {pages} {083046} (\bibinfo {year} {2023}{\natexlab{b}})}\BibitemShut
  {NoStop}%
\bibitem [{\citenamefont {Lier}\ \emph {et~al.}(2023)\citenamefont {Lier},
  \citenamefont {Duclut}, \citenamefont {Bo}, \citenamefont {Armas},
  \citenamefont {J{\"u}licher},\ and\ \citenamefont
  {Sur{\'o}wka}}]{lier2023lift}%
  \BibitemOpen
  \bibfield  {author} {\bibinfo {author} {\bibfnamefont {R.}~\bibnamefont
  {Lier}}, \bibinfo {author} {\bibfnamefont {C.}~\bibnamefont {Duclut}},
  \bibinfo {author} {\bibfnamefont {S.}~\bibnamefont {Bo}}, \bibinfo {author}
  {\bibfnamefont {J.}~\bibnamefont {Armas}}, \bibinfo {author} {\bibfnamefont
  {F.}~\bibnamefont {J{\"u}licher}},\ and\ \bibinfo {author} {\bibfnamefont
  {P.}~\bibnamefont {Sur{\'o}wka}},\ }\bibfield  {title} {\bibinfo {title}
  {Lift force in odd compressible fluids},\ }\href
  {https://doi.org/10.1103/PhysRevE.108.L023101} {\bibfield  {journal}
  {\bibinfo  {journal} {Phys. Rev. E}\ }\textbf {\bibinfo {volume} {108}},\
  \bibinfo {pages} {L023101} (\bibinfo {year} {2023})}\BibitemShut {NoStop}%
\bibitem [{\citenamefont {Daddi-Moussa-Ider}\ \emph
  {et~al.}(2025{\natexlab{a}})\citenamefont {Daddi-Moussa-Ider}, \citenamefont
  {Vilfan},\ and\ \citenamefont {Hosaka}}]{daddi2025analytical}%
  \BibitemOpen
  \bibfield  {author} {\bibinfo {author} {\bibfnamefont {A.}~\bibnamefont
  {Daddi-Moussa-Ider}}, \bibinfo {author} {\bibfnamefont {A.}~\bibnamefont
  {Vilfan}},\ and\ \bibinfo {author} {\bibfnamefont {Y.}~\bibnamefont
  {Hosaka}},\ }\bibfield  {title} {\bibinfo {title} {Analytical solution for
  the hydrodynamic resistance of a disk in a compressible fluid layer with odd
  viscosity on a rigid substrate},\ }\href {https://doi.org/10.1063/5.0249623}
  {\bibfield  {journal} {\bibinfo  {journal} {J. Chem. Phys.}\ }\textbf
  {\bibinfo {volume} {162}},\ \bibinfo {pages} {064103} (\bibinfo {year}
  {2025}{\natexlab{a}})}\BibitemShut {NoStop}%
\bibitem [{\citenamefont {Daddi-Moussa-Ider}\ \emph
  {et~al.}(2025{\natexlab{b}})\citenamefont {Daddi-Moussa-Ider}, \citenamefont
  {Hosaka}, \citenamefont {Tjhung},\ and\ \citenamefont
  {Vilfan}}]{daddi2025hydrodynamic}%
  \BibitemOpen
  \bibfield  {author} {\bibinfo {author} {\bibfnamefont {A.}~\bibnamefont
  {Daddi-Moussa-Ider}}, \bibinfo {author} {\bibfnamefont {Y.}~\bibnamefont
  {Hosaka}}, \bibinfo {author} {\bibfnamefont {E.}~\bibnamefont {Tjhung}},\
  and\ \bibinfo {author} {\bibfnamefont {A.}~\bibnamefont {Vilfan}},\
  }\bibfield  {title} {\bibinfo {title} {Hydrodynamic flow field and frictional
  resistance coefficient of a disk rotating steadily in a compressible fluid
  layer with odd viscosity on a rigid substrate},\ }\href
  {https://doi.org/https://doi.org/10.7566/JPSJ.94.044401} {\bibfield
  {journal} {\bibinfo  {journal} {J. Phys. Soc. Jpn.}\ }\textbf {\bibinfo
  {volume} {94}},\ \bibinfo {pages} {044401} (\bibinfo {year}
  {2025}{\natexlab{b}})}\BibitemShut {NoStop}%
\bibitem [{\citenamefont {Seki}\ \emph {et~al.}(2011)\citenamefont {Seki},
  \citenamefont {Ramachandran},\ and\ \citenamefont
  {Komura}}]{seki2011diffusion}%
  \BibitemOpen
  \bibfield  {author} {\bibinfo {author} {\bibfnamefont {K.}~\bibnamefont
  {Seki}}, \bibinfo {author} {\bibfnamefont {S.}~\bibnamefont {Ramachandran}},\
  and\ \bibinfo {author} {\bibfnamefont {S.}~\bibnamefont {Komura}},\
  }\bibfield  {title} {\bibinfo {title} {Diffusion coefficient of an inclusion
  in a liquid membrane supported by a solvent of arbitrary thickness},\ }\href
  {https://doi.org/https://doi.org/10.1103/PhysRevE.84.021905} {\bibfield
  {journal} {\bibinfo  {journal} {Phys. Rev. E}\ }\textbf {\bibinfo {volume}
  {84}},\ \bibinfo {pages} {021905} (\bibinfo {year} {2011})}\BibitemShut
  {NoStop}%
\bibitem [{\citenamefont {Seki}\ and\ \citenamefont
  {Komura}(1993)}]{seki1993brownian}%
  \BibitemOpen
  \bibfield  {author} {\bibinfo {author} {\bibfnamefont {K.}~\bibnamefont
  {Seki}}\ and\ \bibinfo {author} {\bibfnamefont {S.}~\bibnamefont {Komura}},\
  }\bibfield  {title} {\bibinfo {title} {Brownian dynamics in a thin sheet with
  momentum decay},\ }\href
  {https://doi.org/https://doi.org/10.1103/PhysRevE.47.2377} {\bibfield
  {journal} {\bibinfo  {journal} {Phys. Rev. E}\ }\textbf {\bibinfo {volume}
  {47}},\ \bibinfo {pages} {2377} (\bibinfo {year} {1993})}\BibitemShut
  {NoStop}%
\bibitem [{\citenamefont {Veatch}\ and\ \citenamefont
  {Keller}(2005)}]{veatch2005seeing}%
  \BibitemOpen
  \bibfield  {author} {\bibinfo {author} {\bibfnamefont {S.~L.}\ \bibnamefont
  {Veatch}}\ and\ \bibinfo {author} {\bibfnamefont {S.~L.}\ \bibnamefont
  {Keller}},\ }\bibfield  {title} {\bibinfo {title} {Seeing spots: Complex
  phase behavior in simple membranes},\ }\href
  {https://doi.org/10.1016/j.bbamcr.2005.06.010} {\bibfield  {journal}
  {\bibinfo  {journal} {Biochim. Biophys. Acta Mol. Cell Res.}\ }\textbf
  {\bibinfo {volume} {1746}},\ \bibinfo {pages} {172} (\bibinfo {year}
  {2005})}\BibitemShut {NoStop}%
\bibitem [{\citenamefont {Komura}\ \emph
  {et~al.}(2012{\natexlab{c}})\citenamefont {Komura}, \citenamefont
  {Ramachandran}, \citenamefont {Seki},\ and\ \citenamefont
  {Imai}}]{komura2012dynamics}%
  \BibitemOpen
  \bibfield  {author} {\bibinfo {author} {\bibfnamefont {S.}~\bibnamefont
  {Komura}}, \bibinfo {author} {\bibfnamefont {S.}~\bibnamefont
  {Ramachandran}}, \bibinfo {author} {\bibfnamefont {K.}~\bibnamefont {Seki}},\
  and\ \bibinfo {author} {\bibfnamefont {M.}~\bibnamefont {Imai}},\ }\bibfield
  {title} {\bibinfo {title} {Dynamics of heterogeneity in fluid membranes},\
  }in\ \href
  {https://doi.org/https://doi.org/10.1016/B978-0-12-396534-9.00005-2} {\emph
  {\bibinfo {booktitle} {Advances in Planar Lipid Bilayers and Liposomes}}},\
  Vol.~\bibinfo {volume} {16}\ (\bibinfo  {publisher} {Elsevier},\ \bibinfo
  {year} {2012})\ pp.\ \bibinfo {pages} {129--164}\BibitemShut {NoStop}%
\bibitem [{\citenamefont {Komura}\ and\ \citenamefont
  {Andelman}(2014)}]{komura2014physical}%
  \BibitemOpen
  \bibfield  {author} {\bibinfo {author} {\bibfnamefont {S.}~\bibnamefont
  {Komura}}\ and\ \bibinfo {author} {\bibfnamefont {D.}~\bibnamefont
  {Andelman}},\ }\bibfield  {title} {\bibinfo {title} {Physical aspects of
  heterogeneities in multi-component lipid membranes},\ }\href@noop {}
  {\bibfield  {journal} {\bibinfo  {journal} {Adv. Colloid Interface Sci.}\
  }\textbf {\bibinfo {volume} {208}},\ \bibinfo {pages} {34} (\bibinfo {year}
  {2014})}\BibitemShut {NoStop}%
\bibitem [{\citenamefont {Lingwood}\ and\ \citenamefont
  {Simons}(2010)}]{Lingwood2010}%
  \BibitemOpen
  \bibfield  {author} {\bibinfo {author} {\bibfnamefont {D.}~\bibnamefont
  {Lingwood}}\ and\ \bibinfo {author} {\bibfnamefont {K.}~\bibnamefont
  {Simons}},\ }\bibfield  {title} {\bibinfo {title} {Lipid rafts as a
  membrane-organizing principle},\ }\href
  {https://doi.org/10.1126/science.1174621} {\bibfield  {journal} {\bibinfo
  {journal} {Science}\ }\textbf {\bibinfo {volume} {327}},\ \bibinfo {pages}
  {46} (\bibinfo {year} {2010})}\BibitemShut {NoStop}%
\bibitem [{\citenamefont {Ramachandran}\ \emph
  {et~al.}(2010{\natexlab{a}})\citenamefont {Ramachandran}, \citenamefont
  {Komura}, \citenamefont {Imai},\ and\ \citenamefont
  {Seki}}]{ramachandran2010drag}%
  \BibitemOpen
  \bibfield  {author} {\bibinfo {author} {\bibfnamefont {S.}~\bibnamefont
  {Ramachandran}}, \bibinfo {author} {\bibfnamefont {S.}~\bibnamefont
  {Komura}}, \bibinfo {author} {\bibfnamefont {M.}~\bibnamefont {Imai}},\ and\
  \bibinfo {author} {\bibfnamefont {K.}~\bibnamefont {Seki}},\ }\bibfield
  {title} {\bibinfo {title} {Drag coefficient of a liquid domain in a
  two-dimensional membrane},\ }\href
  {https://doi.org/https://doi.org/10.1140/epje/i2010-10577-3} {\bibfield
  {journal} {\bibinfo  {journal} {Eur. Phys. J. E}\ }\textbf {\bibinfo {volume}
  {31}},\ \bibinfo {pages} {303} (\bibinfo {year}
  {2010}{\natexlab{a}})}\BibitemShut {NoStop}%
\bibitem [{\citenamefont {Dietrich}\ \emph {et~al.}(2001)\citenamefont
  {Dietrich}, \citenamefont {Bagatolli}, \citenamefont {Volovyk}, \citenamefont
  {Thompson}, \citenamefont {Levi}, \citenamefont {Jacobson},\ and\
  \citenamefont {Gratton}}]{Dietrich2001}%
  \BibitemOpen
  \bibfield  {author} {\bibinfo {author} {\bibfnamefont {C.}~\bibnamefont
  {Dietrich}}, \bibinfo {author} {\bibfnamefont {L.~A.}\ \bibnamefont
  {Bagatolli}}, \bibinfo {author} {\bibfnamefont {Z.~N.}\ \bibnamefont
  {Volovyk}}, \bibinfo {author} {\bibfnamefont {N.~L.}\ \bibnamefont
  {Thompson}}, \bibinfo {author} {\bibfnamefont {M.}~\bibnamefont {Levi}},
  \bibinfo {author} {\bibfnamefont {K.}~\bibnamefont {Jacobson}},\ and\
  \bibinfo {author} {\bibfnamefont {E.}~\bibnamefont {Gratton}},\ }\bibfield
  {title} {\bibinfo {title} {Lipid rafts reconstituted in model membranes},\
  }\href {https://doi.org/10.1016/S0006-3495(01)76114-0} {\bibfield  {journal}
  {\bibinfo  {journal} {Biophys. J.}\ }\textbf {\bibinfo {volume} {80}},\
  \bibinfo {pages} {1417} (\bibinfo {year} {2001})}\BibitemShut {NoStop}%
\bibitem [{\citenamefont {Veatch}\ and\ \citenamefont
  {Keller}(2003)}]{Veatch2003}%
  \BibitemOpen
  \bibfield  {author} {\bibinfo {author} {\bibfnamefont {S.~L.}\ \bibnamefont
  {Veatch}}\ and\ \bibinfo {author} {\bibfnamefont {S.~L.}\ \bibnamefont
  {Keller}},\ }\bibfield  {title} {\bibinfo {title} {Separation of liquid
  phases in giant vesicles of ternary mixtures of phospholipids and
  cholesterol},\ }\href {https://doi.org/10.1016/S0006-3495(03)74726-2}
  {\bibfield  {journal} {\bibinfo  {journal} {Biophys. J.}\ }\textbf {\bibinfo
  {volume} {85}},\ \bibinfo {pages} {3074} (\bibinfo {year}
  {2003})}\BibitemShut {NoStop}%
\bibitem [{\citenamefont {Cicuta}\ \emph {et~al.}(2007)\citenamefont {Cicuta},
  \citenamefont {Keller},\ and\ \citenamefont {Veatch}}]{Cicuta2007}%
  \BibitemOpen
  \bibfield  {author} {\bibinfo {author} {\bibfnamefont {P.}~\bibnamefont
  {Cicuta}}, \bibinfo {author} {\bibfnamefont {S.~L.}\ \bibnamefont {Keller}},\
  and\ \bibinfo {author} {\bibfnamefont {S.~L.}\ \bibnamefont {Veatch}},\
  }\bibfield  {title} {\bibinfo {title} {Diffusion of liquid domains in lipid
  bilayer membranes},\ }\href {https://doi.org/10.1021/jp0702088} {\bibfield
  {journal} {\bibinfo  {journal} {J. Phys. Chem. B}\ }\textbf {\bibinfo
  {volume} {111}},\ \bibinfo {pages} {3328} (\bibinfo {year}
  {2007})}\BibitemShut {NoStop}%
\bibitem [{\citenamefont {Or{\"a}dd}\ \emph {et~al.}(2005)\citenamefont
  {Or{\"a}dd}, \citenamefont {Westerman},\ and\ \citenamefont
  {Lindblom}}]{Oradd2005}%
  \BibitemOpen
  \bibfield  {author} {\bibinfo {author} {\bibfnamefont {G.}~\bibnamefont
  {Or{\"a}dd}}, \bibinfo {author} {\bibfnamefont {P.~W.}\ \bibnamefont
  {Westerman}},\ and\ \bibinfo {author} {\bibfnamefont {G.}~\bibnamefont
  {Lindblom}},\ }\bibfield  {title} {\bibinfo {title} {{Lateral diffusion
  coefficients of separate lipid species in a ternary raft-forming bilayer: A
  Pfg-NMR multinuclear study}},\ }\href
  {https://doi.org/10.1529/biophysj.105.061762} {\bibfield  {journal} {\bibinfo
   {journal} {Biophys. J.}\ }\textbf {\bibinfo {volume} {89}},\ \bibinfo
  {pages} {315} (\bibinfo {year} {2005})}\BibitemShut {NoStop}%
\bibitem [{\citenamefont {Baumgart}\ \emph {et~al.}(2003)\citenamefont
  {Baumgart}, \citenamefont {Hess},\ and\ \citenamefont {Webb}}]{Baumgart2003}%
  \BibitemOpen
  \bibfield  {author} {\bibinfo {author} {\bibfnamefont {T.}~\bibnamefont
  {Baumgart}}, \bibinfo {author} {\bibfnamefont {S.~T.}\ \bibnamefont {Hess}},\
  and\ \bibinfo {author} {\bibfnamefont {W.~W.}\ \bibnamefont {Webb}},\
  }\bibfield  {title} {\bibinfo {title} {Imaging coexisting fluid domains in
  biomembrane models coupling curvature and line tension},\ }\href
  {https://doi.org/10.1038/nature02013} {\bibfield  {journal} {\bibinfo
  {journal} {Nature}\ }\textbf {\bibinfo {volume} {425}},\ \bibinfo {pages}
  {821} (\bibinfo {year} {2003})}\BibitemShut {NoStop}%
\bibitem [{\citenamefont {Yanagisawa}\ \emph {et~al.}(2007)\citenamefont
  {Yanagisawa}, \citenamefont {Imai}, \citenamefont {Masui}, \citenamefont
  {Komura},\ and\ \citenamefont {Ohta}}]{Yanagisawa2007}%
  \BibitemOpen
  \bibfield  {author} {\bibinfo {author} {\bibfnamefont {M.}~\bibnamefont
  {Yanagisawa}}, \bibinfo {author} {\bibfnamefont {M.}~\bibnamefont {Imai}},
  \bibinfo {author} {\bibfnamefont {T.}~\bibnamefont {Masui}}, \bibinfo
  {author} {\bibfnamefont {S.}~\bibnamefont {Komura}},\ and\ \bibinfo {author}
  {\bibfnamefont {T.}~\bibnamefont {Ohta}},\ }\bibfield  {title} {\bibinfo
  {title} {Growth dynamics of domains in ternary fluid vesicles},\ }\href
  {https://doi.org/10.1529/biophysj.106.087494} {\bibfield  {journal} {\bibinfo
   {journal} {Biophys. J.}\ }\textbf {\bibinfo {volume} {92}},\ \bibinfo
  {pages} {115} (\bibinfo {year} {2007})}\BibitemShut {NoStop}%
\bibitem [{\citenamefont {Stanich}\ \emph {et~al.}(2013)\citenamefont
  {Stanich}, \citenamefont {Honerkamp-Smith}, \citenamefont {Putzel},
  \citenamefont {Warth}, \citenamefont {Lamprecht}, \citenamefont {Mandal},
  \citenamefont {Mann}, \citenamefont {Hua},\ and\ \citenamefont
  {Keller}}]{Stanich2013}%
  \BibitemOpen
  \bibfield  {author} {\bibinfo {author} {\bibfnamefont {C.~A.}\ \bibnamefont
  {Stanich}}, \bibinfo {author} {\bibfnamefont {A.~R.}\ \bibnamefont
  {Honerkamp-Smith}}, \bibinfo {author} {\bibfnamefont {G.~G.}\ \bibnamefont
  {Putzel}}, \bibinfo {author} {\bibfnamefont {C.~S.}\ \bibnamefont {Warth}},
  \bibinfo {author} {\bibfnamefont {A.~K.}\ \bibnamefont {Lamprecht}}, \bibinfo
  {author} {\bibfnamefont {P.}~\bibnamefont {Mandal}}, \bibinfo {author}
  {\bibfnamefont {E.}~\bibnamefont {Mann}}, \bibinfo {author} {\bibfnamefont
  {T.-A.~D.}\ \bibnamefont {Hua}},\ and\ \bibinfo {author} {\bibfnamefont
  {S.~L.}\ \bibnamefont {Keller}},\ }\bibfield  {title} {\bibinfo {title}
  {Coarsening dynamics of domains in lipid membranes},\ }\href
  {https://doi.org/10.1016/j.bpj.2013.06.013} {\bibfield  {journal} {\bibinfo
  {journal} {Biophys. J.}\ }\textbf {\bibinfo {volume} {105}},\ \bibinfo
  {pages} {444} (\bibinfo {year} {2013})}\BibitemShut {NoStop}%
\bibitem [{\citenamefont {Bray}(2002)}]{Bray2002}%
  \BibitemOpen
  \bibfield  {author} {\bibinfo {author} {\bibfnamefont {A.~J.}\ \bibnamefont
  {Bray}},\ }\bibfield  {title} {\bibinfo {title} {Theory of phase-ordering
  kinetics},\ }\href {https://doi.org/10.1080/00018730110117433} {\bibfield
  {journal} {\bibinfo  {journal} {Adv. Phys.}\ }\textbf {\bibinfo {volume}
  {51}},\ \bibinfo {pages} {481} (\bibinfo {year} {2002})}\BibitemShut
  {NoStop}%
\bibitem [{\citenamefont {Onuki}(2002)}]{Onuki2002}%
  \BibitemOpen
  \bibfield  {author} {\bibinfo {author} {\bibfnamefont {A.}~\bibnamefont
  {Onuki}},\ }\href {https://doi.org/10.1017/CBO9780511535313} {\emph {\bibinfo
  {title} {Phase Transition Dynamics}}}\ (\bibinfo  {publisher} {Cambridge
  University Press},\ \bibinfo {address} {Cambridge},\ \bibinfo {year}
  {2002})\BibitemShut {NoStop}%
\bibitem [{\citenamefont {Ramachandran}\ \emph
  {et~al.}(2010{\natexlab{b}})\citenamefont {Ramachandran}, \citenamefont
  {Komura},\ and\ \citenamefont {Gompper}}]{ramachandran2010effects}%
  \BibitemOpen
  \bibfield  {author} {\bibinfo {author} {\bibfnamefont {S.}~\bibnamefont
  {Ramachandran}}, \bibinfo {author} {\bibfnamefont {S.}~\bibnamefont
  {Komura}},\ and\ \bibinfo {author} {\bibfnamefont {G.}~\bibnamefont
  {Gompper}},\ }\bibfield  {title} {\bibinfo {title} {Effects of an embedding
  bulk fluid on phase separation dynamics in a thin liquid film},\ }\href
  {https://doi.org/10.1209/0295-5075/89/56001} {\bibfield  {journal} {\bibinfo
  {journal} {EPL}\ }\textbf {\bibinfo {volume} {89}},\ \bibinfo {pages} {56001}
  (\bibinfo {year} {2010}{\natexlab{b}})}\BibitemShut {NoStop}%
\bibitem [{\citenamefont {Veatch}\ \emph {et~al.}(2007)\citenamefont {Veatch},
  \citenamefont {Soubias}, \citenamefont {Keller},\ and\ \citenamefont
  {Gawrisch}}]{Veatch2007Critical}%
  \BibitemOpen
  \bibfield  {author} {\bibinfo {author} {\bibfnamefont {S.~L.}\ \bibnamefont
  {Veatch}}, \bibinfo {author} {\bibfnamefont {O.}~\bibnamefont {Soubias}},
  \bibinfo {author} {\bibfnamefont {S.~L.}\ \bibnamefont {Keller}},\ and\
  \bibinfo {author} {\bibfnamefont {K.}~\bibnamefont {Gawrisch}},\ }\bibfield
  {title} {\bibinfo {title} {Critical fluctuations in domain-forming lipid
  mixtures},\ }\href {https://doi.org/10.1073/pnas.0703513104} {\bibfield
  {journal} {\bibinfo  {journal} {Proc. Natl. Acad. Sci. U.S.A.}\ }\textbf
  {\bibinfo {volume} {104}},\ \bibinfo {pages} {17650} (\bibinfo {year}
  {2007})}\BibitemShut {NoStop}%
\bibitem [{\citenamefont {Honerkamp-Smith}\ \emph {et~al.}(2008)\citenamefont
  {Honerkamp-Smith}, \citenamefont {Cicuta}, \citenamefont {Collins},
  \citenamefont {Veatch}, \citenamefont {den Nijs}, \citenamefont {Schick},\
  and\ \citenamefont {Keller}}]{HonerkampSmith2008}%
  \BibitemOpen
  \bibfield  {author} {\bibinfo {author} {\bibfnamefont {A.~R.}\ \bibnamefont
  {Honerkamp-Smith}}, \bibinfo {author} {\bibfnamefont {P.}~\bibnamefont
  {Cicuta}}, \bibinfo {author} {\bibfnamefont {M.~D.}\ \bibnamefont {Collins}},
  \bibinfo {author} {\bibfnamefont {S.~L.}\ \bibnamefont {Veatch}}, \bibinfo
  {author} {\bibfnamefont {M.}~\bibnamefont {den Nijs}}, \bibinfo {author}
  {\bibfnamefont {M.}~\bibnamefont {Schick}},\ and\ \bibinfo {author}
  {\bibfnamefont {S.~L.}\ \bibnamefont {Keller}},\ }\bibfield  {title}
  {\bibinfo {title} {Line tensions, correlation lengths, and critical exponents
  in lipid membranes near critical points},\ }\href
  {https://doi.org/10.1529/biophysj.107.128421} {\bibfield  {journal} {\bibinfo
   {journal} {Biophys. J.}\ }\textbf {\bibinfo {volume} {95}},\ \bibinfo
  {pages} {236} (\bibinfo {year} {2008})}\BibitemShut {NoStop}%
\bibitem [{\citenamefont {Veatch}\ \emph {et~al.}(2008)\citenamefont {Veatch},
  \citenamefont {Cicuta}, \citenamefont {Sengupta}, \citenamefont
  {Honerkamp-Smith}, \citenamefont {Holowka},\ and\ \citenamefont
  {Baird}}]{Veatch2008GPMV}%
  \BibitemOpen
  \bibfield  {author} {\bibinfo {author} {\bibfnamefont {S.~L.}\ \bibnamefont
  {Veatch}}, \bibinfo {author} {\bibfnamefont {P.}~\bibnamefont {Cicuta}},
  \bibinfo {author} {\bibfnamefont {P.}~\bibnamefont {Sengupta}}, \bibinfo
  {author} {\bibfnamefont {A.}~\bibnamefont {Honerkamp-Smith}}, \bibinfo
  {author} {\bibfnamefont {D.}~\bibnamefont {Holowka}},\ and\ \bibinfo {author}
  {\bibfnamefont {B.}~\bibnamefont {Baird}},\ }\bibfield  {title} {\bibinfo
  {title} {Critical fluctuations in plasma membrane vesicles},\ }\href
  {https://doi.org/10.1021/cb800012x} {\bibfield  {journal} {\bibinfo
  {journal} {ACS Chem. Biol.}\ }\textbf {\bibinfo {volume} {3}},\ \bibinfo
  {pages} {287} (\bibinfo {year} {2008})}\BibitemShut {NoStop}%
\bibitem [{\citenamefont {Seki}\ \emph {et~al.}(2007)\citenamefont {Seki},
  \citenamefont {Komura},\ and\ \citenamefont {Imai}}]{seki2007concentration}%
  \BibitemOpen
  \bibfield  {author} {\bibinfo {author} {\bibfnamefont {K.}~\bibnamefont
  {Seki}}, \bibinfo {author} {\bibfnamefont {S.}~\bibnamefont {Komura}},\ and\
  \bibinfo {author} {\bibfnamefont {M.}~\bibnamefont {Imai}},\ }\bibfield
  {title} {\bibinfo {title} {Concentration fluctuations in binary fluid
  membranes},\ }\href {https://doi.org/10.1088/0953-8984/19/7/072101}
  {\bibfield  {journal} {\bibinfo  {journal} {J. Phys.: Condens. Matter}\
  }\textbf {\bibinfo {volume} {19}},\ \bibinfo {pages} {072101} (\bibinfo
  {year} {2007})}\BibitemShut {NoStop}%
\bibitem [{\citenamefont {Ramachandran}\ \emph
  {et~al.}(2011{\natexlab{a}})\citenamefont {Ramachandran}, \citenamefont
  {Komura}, \citenamefont {Seki},\ and\ \citenamefont
  {Imai}}]{ramachandran2011hydrodynamic}%
  \BibitemOpen
  \bibfield  {author} {\bibinfo {author} {\bibfnamefont {S.}~\bibnamefont
  {Ramachandran}}, \bibinfo {author} {\bibfnamefont {S.}~\bibnamefont
  {Komura}}, \bibinfo {author} {\bibfnamefont {K.}~\bibnamefont {Seki}},\ and\
  \bibinfo {author} {\bibfnamefont {M.}~\bibnamefont {Imai}},\ }\bibfield
  {title} {\bibinfo {title} {Hydrodynamic effects on concentration fluctuations
  in multicomponent membranes},\ }\href
  {https://doi.org/https://doi.org/10.1039/C0SM00783H} {\bibfield  {journal}
  {\bibinfo  {journal} {Soft Matter}\ }\textbf {\bibinfo {volume} {7}},\
  \bibinfo {pages} {1524} (\bibinfo {year} {2011}{\natexlab{a}})}\BibitemShut
  {NoStop}%
\bibitem [{\citenamefont {Lauga}(2020)}]{lauga2020fluid}%
  \BibitemOpen
  \bibfield  {author} {\bibinfo {author} {\bibfnamefont {E.}~\bibnamefont
  {Lauga}},\ }\href {https://doi.org/https://doi.org/10.1017/9781316796047}
  {\emph {\bibinfo {title} {The Fluid Dynamics of Cell Motility}}},\
  Vol.~\bibinfo {volume} {62}\ (\bibinfo  {publisher} {Cambridge University
  Press},\ \bibinfo {year} {2020})\BibitemShut {NoStop}%
\bibitem [{\citenamefont {Pozrikidis}(1992)}]{pozrikidis1992}%
  \BibitemOpen
  \bibfield  {author} {\bibinfo {author} {\bibfnamefont {C.}~\bibnamefont
  {Pozrikidis}},\ }\href@noop {} {\emph {\bibinfo {title} {Boundary Integral
  and Singularity Methods for Linearized Viscous Flow}}}\ (\bibinfo
  {publisher} {Cambridge University Press},\ \bibinfo {address} {Cambridge,
  England},\ \bibinfo {year} {1992})\BibitemShut {NoStop}%
\bibitem [{\citenamefont {Ramachandran}\ \emph
  {et~al.}(2011{\natexlab{b}})\citenamefont {Ramachandran}, \citenamefont
  {Komura}, \citenamefont {Seki},\ and\ \citenamefont
  {Gompper}}]{ramachandran2011}%
  \BibitemOpen
  \bibfield  {author} {\bibinfo {author} {\bibfnamefont {S.}~\bibnamefont
  {Ramachandran}}, \bibinfo {author} {\bibfnamefont {S.}~\bibnamefont
  {Komura}}, \bibinfo {author} {\bibfnamefont {K.}~\bibnamefont {Seki}},\ and\
  \bibinfo {author} {\bibfnamefont {G.}~\bibnamefont {Gompper}},\ }\bibfield
  {title} {\bibinfo {title} {Dynamics of a polymer chain confined in a
  membrane},\ }\href
  {https://doi.org/https://doi.org/10.1140/epje/i2011-11046-3} {\bibfield
  {journal} {\bibinfo  {journal} {Eur. Phys. J. E}\ }\textbf {\bibinfo {volume}
  {34}},\ \bibinfo {pages} {46} (\bibinfo {year}
  {2011}{\natexlab{b}})}\BibitemShut {NoStop}%
\bibitem [{\citenamefont {Oppenheimer}\ and\ \citenamefont
  {Diamant}(2010)}]{oppenheimer2010}%
  \BibitemOpen
  \bibfield  {author} {\bibinfo {author} {\bibfnamefont {N.}~\bibnamefont
  {Oppenheimer}}\ and\ \bibinfo {author} {\bibfnamefont {H.}~\bibnamefont
  {Diamant}},\ }\bibfield  {title} {\bibinfo {title} {Correlated dynamics of
  inclusions in a supported membrane},\ }\href
  {https://doi.org/https://doi.org/10.1103/PhysRevE.82.041912} {\bibfield
  {journal} {\bibinfo  {journal} {Phys. Rev. E}\ }\textbf {\bibinfo {volume}
  {82}},\ \bibinfo {pages} {041912} (\bibinfo {year} {2010})}\BibitemShut
  {NoStop}%
\bibitem [{\citenamefont {Komura}\ and\ \citenamefont
  {Seki}(1995)}]{komura1995diffusion}%
  \BibitemOpen
  \bibfield  {author} {\bibinfo {author} {\bibfnamefont {S.}~\bibnamefont
  {Komura}}\ and\ \bibinfo {author} {\bibfnamefont {K.}~\bibnamefont {Seki}},\
  }\bibfield  {title} {\bibinfo {title} {Diffusion constant of a polymer chain
  in biomembranes},\ }\href
  {https://doi.org/https://doi.org/10.1051/jp2:1995109} {\bibfield  {journal}
  {\bibinfo  {journal} {J. Phys. II France}\ }\textbf {\bibinfo {volume} {5}},\
  \bibinfo {pages} {5} (\bibinfo {year} {1995})}\BibitemShut {NoStop}%
\bibitem [{\citenamefont {Mikhailov}\ and\ \citenamefont
  {Kapral}(2015)}]{mikhailov2015}%
  \BibitemOpen
  \bibfield  {author} {\bibinfo {author} {\bibfnamefont {A.~S.}\ \bibnamefont
  {Mikhailov}}\ and\ \bibinfo {author} {\bibfnamefont {R.}~\bibnamefont
  {Kapral}},\ }\bibfield  {title} {\bibinfo {title} {{Hydrodynamic collective
  effects of active protein machines in solution and lipid bilayers}},\ }\href
  {https://doi.org/10.1073/pnas.1506825112} {\bibfield  {journal} {\bibinfo
  {journal} {Proc. Natl. Acad. Sci. U.S.A.}\ }\textbf {\bibinfo {volume}
  {112}},\ \bibinfo {pages} {E3639} (\bibinfo {year} {2015})}\BibitemShut
  {NoStop}%
\bibitem [{\citenamefont {Hosaka}\ \emph {et~al.}(2017)\citenamefont {Hosaka},
  \citenamefont {Yasuda}, \citenamefont {Okamoto},\ and\ \citenamefont
  {Komura}}]{hosaka2017}%
  \BibitemOpen
  \bibfield  {author} {\bibinfo {author} {\bibfnamefont {Y.}~\bibnamefont
  {Hosaka}}, \bibinfo {author} {\bibfnamefont {K.}~\bibnamefont {Yasuda}},
  \bibinfo {author} {\bibfnamefont {R.}~\bibnamefont {Okamoto}},\ and\ \bibinfo
  {author} {\bibfnamefont {S.}~\bibnamefont {Komura}},\ }\bibfield  {title}
  {\bibinfo {title} {Lateral diffusion induced by active proteins in a
  biomembrane},\ }\href {https://doi.org/10.1103/PhysRevE.95.052407} {\bibfield
   {journal} {\bibinfo  {journal} {Phys. Rev. E}\ }\textbf {\bibinfo {volume}
  {95}},\ \bibinfo {pages} {052407} (\bibinfo {year} {2017})}\BibitemShut
  {NoStop}%
\bibitem [{\citenamefont {Manikantan}(2020)}]{manikantan2020}%
  \BibitemOpen
  \bibfield  {author} {\bibinfo {author} {\bibfnamefont {H.}~\bibnamefont
  {Manikantan}},\ }\bibfield  {title} {\bibinfo {title} {Tunable collective
  dynamics of active inclusions in viscous membranes},\ }\href
  {https://doi.org/https://doi.org/10.1103/PhysRevLett.125.268101} {\bibfield
  {journal} {\bibinfo  {journal} {Phys. Rev. Lett.}\ }\textbf {\bibinfo
  {volume} {125}},\ \bibinfo {pages} {268101} (\bibinfo {year}
  {2020})}\BibitemShut {NoStop}%
\bibitem [{\citenamefont {Purcell}(2014)}]{purcell2014life}%
  \BibitemOpen
  \bibfield  {author} {\bibinfo {author} {\bibfnamefont {E.~M.}\ \bibnamefont
  {Purcell}},\ }\bibfield  {title} {\bibinfo {title} {Life at low {R}eynolds
  number},\ }in\ \href@noop {} {\emph {\bibinfo {booktitle} {Physics and our
  world: reissue of the proceedings of a symposium in honor of Victor F
  Weisskopf}}}\ (\bibinfo {organization} {World Scientific},\ \bibinfo {year}
  {2014})\ pp.\ \bibinfo {pages} {47--67}\BibitemShut {NoStop}%
\bibitem [{\citenamefont {Ota}\ \emph {et~al.}(2018)\citenamefont {Ota},
  \citenamefont {Hosaka}, \citenamefont {Yasuda},\ and\ \citenamefont
  {Komura}}]{ota2018three}%
  \BibitemOpen
  \bibfield  {author} {\bibinfo {author} {\bibfnamefont {Y.}~\bibnamefont
  {Ota}}, \bibinfo {author} {\bibfnamefont {Y.}~\bibnamefont {Hosaka}},
  \bibinfo {author} {\bibfnamefont {K.}~\bibnamefont {Yasuda}},\ and\ \bibinfo
  {author} {\bibfnamefont {S.}~\bibnamefont {Komura}},\ }\bibfield  {title}
  {\bibinfo {title} {Three-disk microswimmer in a supported fluid membrane},\
  }\href {https://doi.org/https://doi.org/10.1103/PhysRevE.97.052612}
  {\bibfield  {journal} {\bibinfo  {journal} {Phys. Rev. E}\ }\textbf {\bibinfo
  {volume} {97}},\ \bibinfo {pages} {052612} (\bibinfo {year}
  {2018})}\BibitemShut {NoStop}%
\bibitem [{\citenamefont {Najafi}\ and\ \citenamefont
  {Golestanian}(2004)}]{najafi2004}%
  \BibitemOpen
  \bibfield  {author} {\bibinfo {author} {\bibfnamefont {A.}~\bibnamefont
  {Najafi}}\ and\ \bibinfo {author} {\bibfnamefont {R.}~\bibnamefont
  {Golestanian}},\ }\bibfield  {title} {\bibinfo {title} {Simple swimmer at low
  {Reynolds} number: Three linked spheres},\ }\href
  {https://doi.org/10.1103/PhysRevE.69.062901} {\bibfield  {journal} {\bibinfo
  {journal} {Phys. Rev. E}\ }\textbf {\bibinfo {volume} {69}},\ \bibinfo
  {pages} {062901} (\bibinfo {year} {2004})}\BibitemShut {NoStop}%
\bibitem [{\citenamefont {Golestanian}\ and\ \citenamefont
  {Ajdari}(2008)}]{golestanian2008}%
  \BibitemOpen
  \bibfield  {author} {\bibinfo {author} {\bibfnamefont {R.}~\bibnamefont
  {Golestanian}}\ and\ \bibinfo {author} {\bibfnamefont {A.}~\bibnamefont
  {Ajdari}},\ }\bibfield  {title} {\bibinfo {title} {Analytic results for the
  three-sphere swimmer at low {R}eynolds number},\ }\href
  {https://doi.org/10.1103/PhysRevE.77.036308} {\bibfield  {journal} {\bibinfo
  {journal} {Phys. Rev. E}\ }\textbf {\bibinfo {volume} {77}},\ \bibinfo
  {pages} {036308} (\bibinfo {year} {2008})}\BibitemShut {NoStop}%
\bibitem [{\citenamefont {Hosaka}\ and\ \citenamefont
  {Komura}(2022)}]{hosaka2022nonequilibrium}%
  \BibitemOpen
  \bibfield  {author} {\bibinfo {author} {\bibfnamefont {Y.}~\bibnamefont
  {Hosaka}}\ and\ \bibinfo {author} {\bibfnamefont {S.}~\bibnamefont
  {Komura}},\ }\bibfield  {title} {\bibinfo {title} {Nonequilibrium transport
  induced by biological nanomachines},\ }\href
  {https://doi.org/https://doi.org/10.1142/S1793048022310026} {\bibfield
  {journal} {\bibinfo  {journal} {Biophys. Rev. Lett.}\ }\textbf {\bibinfo
  {volume} {17}},\ \bibinfo {pages} {51} (\bibinfo {year} {2022})}\BibitemShut
  {NoStop}%
\bibitem [{\citenamefont {Banerjee}\ \emph {et~al.}(2017)\citenamefont
  {Banerjee}, \citenamefont {Souslov}, \citenamefont {Abanov},\ and\
  \citenamefont {Vitelli}}]{banerjee2017}%
  \BibitemOpen
  \bibfield  {author} {\bibinfo {author} {\bibfnamefont {D.}~\bibnamefont
  {Banerjee}}, \bibinfo {author} {\bibfnamefont {A.}~\bibnamefont {Souslov}},
  \bibinfo {author} {\bibfnamefont {A.~G.}\ \bibnamefont {Abanov}},\ and\
  \bibinfo {author} {\bibfnamefont {V.}~\bibnamefont {Vitelli}},\ }\bibfield
  {title} {\bibinfo {title} {Odd viscosity in chiral active fluids},\ }\href
  {https://doi.org/10.1038/s41467-017-01378-7} {\bibfield  {journal} {\bibinfo
  {journal} {Nat. Commun.}\ }\textbf {\bibinfo {volume} {8}},\ \bibinfo {pages}
  {1573} (\bibinfo {year} {2017})}\BibitemShut {NoStop}%
\bibitem [{\citenamefont {Avron}(1998)}]{avron1998}%
  \BibitemOpen
  \bibfield  {author} {\bibinfo {author} {\bibfnamefont {J.~E.}\ \bibnamefont
  {Avron}},\ }\bibfield  {title} {\bibinfo {title} {Odd viscosity},\ }\href
  {https://doi.org/10.1023/A:1023084404080} {\bibfield  {journal} {\bibinfo
  {journal} {J. Stat. Phys.}\ }\textbf {\bibinfo {volume} {92}},\ \bibinfo
  {pages} {543} (\bibinfo {year} {1998})}\BibitemShut {NoStop}%
\bibitem [{\citenamefont {Ganeshan}\ and\ \citenamefont
  {Abanov}(2017)}]{ganeshan2017}%
  \BibitemOpen
  \bibfield  {author} {\bibinfo {author} {\bibfnamefont {S.}~\bibnamefont
  {Ganeshan}}\ and\ \bibinfo {author} {\bibfnamefont {A.~G.}\ \bibnamefont
  {Abanov}},\ }\bibfield  {title} {\bibinfo {title} {Odd viscosity in
  two-dimensional incompressible fluids},\ }\href
  {https://doi.org/10.1103/PhysRevFluids.2.094101} {\bibfield  {journal}
  {\bibinfo  {journal} {Phys. Rev. Fluids}\ }\textbf {\bibinfo {volume} {2}},\
  \bibinfo {pages} {094101} (\bibinfo {year} {2017})}\BibitemShut {NoStop}%
\bibitem [{\citenamefont {Hosaka}\ \emph
  {et~al.}(2021{\natexlab{b}})\citenamefont {Hosaka}, \citenamefont {Komura},\
  and\ \citenamefont {Andelman}}]{hosaka2021hydrodynamic}%
  \BibitemOpen
  \bibfield  {author} {\bibinfo {author} {\bibfnamefont {Y.}~\bibnamefont
  {Hosaka}}, \bibinfo {author} {\bibfnamefont {S.}~\bibnamefont {Komura}},\
  and\ \bibinfo {author} {\bibfnamefont {D.}~\bibnamefont {Andelman}},\
  }\bibfield  {title} {\bibinfo {title} {Hydrodynamic lift of a two-dimensional
  liquid domain with odd viscosity},\ }\href
  {https://doi.org/10.1103/PhysRevE.104.064613} {\bibfield  {journal} {\bibinfo
   {journal} {Phys. Rev. E}\ }\textbf {\bibinfo {volume} {104}},\ \bibinfo
  {pages} {064613} (\bibinfo {year} {2021}{\natexlab{b}})}\BibitemShut
  {NoStop}%
\bibitem [{\citenamefont {Masoud}\ and\ \citenamefont
  {Stone}(2019)}]{masoud2019}%
  \BibitemOpen
  \bibfield  {author} {\bibinfo {author} {\bibfnamefont {H.}~\bibnamefont
  {Masoud}}\ and\ \bibinfo {author} {\bibfnamefont {H.~A.}\ \bibnamefont
  {Stone}},\ }\bibfield  {title} {\bibinfo {title} {The reciprocal theorem in
  fluid dynamics and transport phenomena},\ }\href
  {https://doi.org/10.1017/jfm.2019.553} {\bibfield  {journal} {\bibinfo
  {journal} {J. Fluid Mech.}\ }\textbf {\bibinfo {volume} {879}},\ \bibinfo
  {pages} {P1} (\bibinfo {year} {2019})}\BibitemShut {NoStop}%
\bibitem [{\citenamefont {Fruchart}\ \emph {et~al.}(2023)\citenamefont
  {Fruchart}, \citenamefont {Scheibner},\ and\ \citenamefont
  {Vitelli}}]{fruchart2023odd}%
  \BibitemOpen
  \bibfield  {author} {\bibinfo {author} {\bibfnamefont {M.}~\bibnamefont
  {Fruchart}}, \bibinfo {author} {\bibfnamefont {C.}~\bibnamefont
  {Scheibner}},\ and\ \bibinfo {author} {\bibfnamefont {V.}~\bibnamefont
  {Vitelli}},\ }\bibfield  {title} {\bibinfo {title} {Odd viscosity and odd
  elasticity},\ }\href
  {https://doi.org/10.1146/annurev-conmatphys-040821-125506} {\bibfield
  {journal} {\bibinfo  {journal} {Annu. Rev. Condens. Matter Phys.}\ }\textbf
  {\bibinfo {volume} {14}},\ \bibinfo {pages} {471} (\bibinfo {year}
  {2023})}\BibitemShut {NoStop}%
\bibitem [{\citenamefont {Hosaka}\ \emph
  {et~al.}(2023{\natexlab{c}})\citenamefont {Hosaka}, \citenamefont
  {Golestanian},\ and\ \citenamefont {Vilfan}}]{hosaka2023lorentz}%
  \BibitemOpen
  \bibfield  {author} {\bibinfo {author} {\bibfnamefont {Y.}~\bibnamefont
  {Hosaka}}, \bibinfo {author} {\bibfnamefont {R.}~\bibnamefont
  {Golestanian}},\ and\ \bibinfo {author} {\bibfnamefont {A.}~\bibnamefont
  {Vilfan}},\ }\bibfield  {title} {\bibinfo {title} {Lorentz reciprocal theorem
  in fluids with odd viscosity},\ }\href
  {https://doi.org/10.1103/PhysRevLett.131.178303} {\bibfield  {journal}
  {\bibinfo  {journal} {Phys. Rev. Lett.}\ }\textbf {\bibinfo {volume} {131}},\
  \bibinfo {pages} {178303} (\bibinfo {year} {2023}{\natexlab{c}})}\BibitemShut
  {NoStop}%
\bibitem [{\citenamefont {Chandran~Suja}\ \emph {et~al.}(2025)\citenamefont
  {Chandran~Suja}, \citenamefont {Oppenheimer},\ and\ \citenamefont
  {Stone}}]{chandran2025hydrodynamics}%
  \BibitemOpen
  \bibfield  {author} {\bibinfo {author} {\bibfnamefont {V.}~\bibnamefont
  {Chandran~Suja}}, \bibinfo {author} {\bibfnamefont {N.}~\bibnamefont
  {Oppenheimer}},\ and\ \bibinfo {author} {\bibfnamefont {H.~A.}\ \bibnamefont
  {Stone}},\ }\bibfield  {title} {\bibinfo {title} {Hydrodynamics of molecular
  rotors in lipid membranes},\ }\href
  {https://doi.org/https://doi.org/10.1103/PhysRevFluids.10.L041101} {\bibfield
   {journal} {\bibinfo  {journal} {Phys. Rev. Fluids}\ }\textbf {\bibinfo
  {volume} {10}},\ \bibinfo {pages} {L041101} (\bibinfo {year}
  {2025})}\BibitemShut {NoStop}%
\bibitem [{\citenamefont {Levine}\ \emph {et~al.}(2004)\citenamefont {Levine},
  \citenamefont {Liverpool},\ and\ \citenamefont {MacKintosh}}]{Levine2004}%
  \BibitemOpen
  \bibfield  {author} {\bibinfo {author} {\bibfnamefont {A.~J.}\ \bibnamefont
  {Levine}}, \bibinfo {author} {\bibfnamefont {T.~B.}\ \bibnamefont
  {Liverpool}},\ and\ \bibinfo {author} {\bibfnamefont {F.~C.}\ \bibnamefont
  {MacKintosh}},\ }\bibfield  {title} {\bibinfo {title} {Dynamics of rigid and
  flexible extended bodies in viscous films and membranes},\ }\href
  {https://doi.org/10.1103/PhysRevLett.93.038102} {\bibfield  {journal}
  {\bibinfo  {journal} {Phys. Rev. Lett.}\ }\textbf {\bibinfo {volume} {93}},\
  \bibinfo {pages} {038102} (\bibinfo {year} {2004})}\BibitemShut {NoStop}%
\bibitem [{\citenamefont {Fischer}(2004)}]{Fischer2004}%
  \BibitemOpen
  \bibfield  {author} {\bibinfo {author} {\bibfnamefont {T.~M.}\ \bibnamefont
  {Fischer}},\ }\bibfield  {title} {\bibinfo {title} {The drag on needles
  moving in a {Langmuir} monolayer},\ }\href
  {https://doi.org/10.1017/S0022112003006608} {\bibfield  {journal} {\bibinfo
  {journal} {J. Fluid Mech.}\ }\textbf {\bibinfo {volume} {498}},\ \bibinfo
  {pages} {123} (\bibinfo {year} {2004})}\BibitemShut {NoStop}%
\bibitem [{\citenamefont {Manikantan}(2024)}]{Manikantan_2024}%
  \BibitemOpen
  \bibfield  {author} {\bibinfo {author} {\bibfnamefont {H.}~\bibnamefont
  {Manikantan}},\ }\bibfield  {title} {\bibinfo {title} {Stability of a
  dispersion of elongated particles embedded in a viscous membrane},\ }\href
  {https://doi.org/10.1017/jfm.2024.395} {\bibfield  {journal} {\bibinfo
  {journal} {J. Fluid Mech.}\ }\textbf {\bibinfo {volume} {987}},\ \bibinfo
  {pages} {R4} (\bibinfo {year} {2024})}\BibitemShut {NoStop}%
\end{thebibliography}

\end{document}